\documentclass[showpacs,aps,prd,reprint,groupedaddress,superscriptaddress,preprintnumbers,floatfix,a4paper,nofootinbib,amsmath,amssymb,longbibliography]{revtex4-1}
\usepackage{graphicx}
\usepackage{booktabs}
\usepackage[hidelinks]{hyperref}
\usepackage{color}
\hypersetup{
    colorlinks,
    linkcolor={blue},
    citecolor={blue},
    urlcolor={blue}
}
\usepackage{multirow} 
\usepackage{amssymb,amsmath,amsfonts}
\usepackage[utf8]{inputenc}
\usepackage{nicefrac}
\usepackage{braket}
\usepackage{slashed}
\usepackage{float}
\usepackage{siunitx}
\def\chitop{\chi_\mathrm{top}}

\def\MSbar{$\overline{\mathrm{MS}}$}
\def\gflow{g^2_{\mathrm{flow}}}
\def\gfit{g^2_{\mathrm{fit}}}
\def\mubar{\overline{\mu}}
\def\tauf{\tau_{\mathrm{f}}}
\def\taufa{\tau_{\mathrm{f}1}}
\def\taufb{\tau_{\mathrm{f}2}}
\def\taufi{\tau_{\mathrm{f}i}}
\def\Eq#1{Eq.~(\ref{#1})}
\def\Tr{\,\mathrm{Tr}\:}
\def\Rlatt{R_{\mathrm{latt}}}

\let\originalleft\left
\let\originalright\right
\renewcommand{\left}{\mathopen{}\mathclose\bgroup\originalleft}
\renewcommand{\right}{\aftergroup\egroup\originalright}

\def\la{\langle}
\def\ra{\rangle}

\begin{document}
	
\title{\boldmath
Step scaling with gradient flow and finite temperature
}
	
\author{Parikshit M. Junnarkar}
\email{parikshit@theorie.ikp.physik.tu-darmstadt.de}
\author{Guy D. Moore}
\email{guymoore@theorie.ikp.physik.tu-darmstadt.de}
\author{Aidan Chaumet}
\email{achaumet@theorie.ikp.physik.tu-darmstadt.de}
\affiliation{Institut f\"ur Kernphysik (Theoriezentrum), Technische Universit\"at Darmstadt,\\ Schlossgartenstra{\ss}e 2, D-64289 Darmstadt, Germany.}

\begin{abstract}
We combine gradient flow, step-scaling, and finite-temperature boundary conditions to scale-set 2+1+1 flavor QCD lattices with physical HISQ quarks at multiple spacings down to $a=0.01378\,\mathrm{fm}$, such that they represent the same temperature at the percent level and the same quark mass to a few percent.
This preparatory work will allow the evaluation and continuum extrapolation of the topological susceptibility at up to 1 GeV temperatures with good control over quark-mass effects.
\end{abstract}
	
\date{\today}
\maketitle

\section{Introduction}
This paper will be about the technical problem of scale-setting on the lattice at small lattice spacings and with special attention to precise quark mass determination.
But we will start the introduction with an explanation of what, scientifically, we want such scale setting for, and in particular why it is important to perform the fermion-mass matching accurately for our application.

The axion is a hypothetical particle in Standard Model (SM) extensions which solve the strong CP problem
\cite{Weinberg:1977ma,Wilczek:1977pj}.
In addition, the axion provides a candidate for the Dark Matter of the Universe \cite{Preskill:1982cy,Abbott:1982af,Dine:1982ah}, thereby solving both a field theory puzzle and a cosmology puzzle within a single simple SM extension.
It would be valuable to be able to make a concrete prediction for the relation between the axion mass and the dark matter axion number density.
If the axion field is made uniform by inflation, this requires knowledge of its initial angle; however the lack of large isocurvature fluctuations strongly constrains this case \cite{Visinelli:2009zm}.
On the other hand, post-inflation physics can give the axion spatially random initial conditions, leading to complex network dynamics but eventually to a precise relation between the axion mass and dark matter density
\cite{Klaer:2017qhr,Vaquero:2018tib,Buschmann:2019icd}.
However, to make this connection precise, we need (among other things) a rather precise determination of the topological susceptibility of Quantum Chromodynamics in a range of temperatures from approximately 400 to 1100 MeV temperature
\cite{Klaer:2017qhr,Moore:2017ond}.

Several lattice studies have investigated the topological susceptibility of full QCD at high temperatures
\cite{Petreczky:2016vrs,Bonati:2018blm,Athenodorou:2022aay},
but only one such study reaches 1000 MeV or higher
\cite{Borsanyi:2016ksw}, and this study uses an indirect method.
A second study with independent techniques would be valuable.
Our goal is to carry out such a study using the reweighting techniques of Jahn et al \cite{Jahn:2018dke,Jahn:2020oqf}.
This will require multiple lattices with temporal extents of $N_\tau = 8$ to 14 (to enable a continuum extrapolation) at temperatures from around 400 to over 1000 MeV, using 2+1+1 flavor ensembles.
The susceptibility is a very strong power of the temperature, with $\chi \propto T^{-8}$ for 3-flavor QCD at lowest order in perturbation theory \cite{Gross:1980br}.  Therefore the dimensionless susceptibility which one directly measures on the lattice, $\chi a^4$, scales as 
$\chi a^4 \propto a^{12}$,
and errors in the lattice spacing are amplified by a factor of about 12 in the determined susceptibility.
Because the high-temperature topological susceptibility is proportional to a product of light quark masses \cite{Gross:1980br,Kanazawa:2014cua}, any error in the quark mass determination similarly leads to large errors in the determined susceptibility.
Therefore our goal will be to find a set of lattice parameters which represent the correct physical point at a set of lattice spacings representing 3 temperatures between 400 and 1100 MeV, each with 4 spacings with $N_\tau = 8, 10, 12, 14$.
And our precision goal will be 1 to 2\% in the lattice spacing and at most a few percent in the quark masses.
Because instantons have associated zero modes, it is important to use a fermionic implementation which does a good job handling chirality, so we want the scale setting within the HISQ framework \cite{Follana:2006rc}.

Scale setting is one of the most important calculations in Lattice QCD since the precision of scale determination sets the precision of any computed observable.
There are two approaches.
One, direct scale setting, involves measuring one vacuum observable per parameter of the theory.
For instance, one can use pseudoscalar masses to establish each quark mass and the Sommer parameter $r_0$
\cite{Sommer:1993ce},
a pseudoscalar decay constant
or a baryon mass such as the $\Omega$ mass
as an additional physical scale to fit the coupling strength.
The finest lattice spacing at which such a direct scale setting has been carried out for 2+1+1 flavor HISQ fermions is $a = 0.03215\,\mathrm{fm}$
\cite{MILC:2012znn,Bazavov:2017lyh}.
This approach becomes rapidly more challenging as one goes to smaller lattice spacing, and since we need more than a factor of 2 finer lattices than this finest known HISQ point, this approach will not be practical for us.

The alternative is step scaling \cite{Luscher:1991wu},
in which we start with a lattice where the scale and quark masses have already been established, and determine the scales for another lattice as a ratio with respect to these known values.
The matching involved is essentially a UV matching problem, and is insensitive to infrared affects.
Therefore one can introduce an IR regularization which eliminates topology, fermionic zero modes, and long correlation lengths which would otherwise impede numerical efficiency.
This is true so long as the regularization is identical for all lattices.
So for instance, using a finite lattice volume is acceptable as long as the physical lattice volume is the same for all systems which are being compared.
In the most modern implementation of this approach
\cite{Nada:2020jay}, one uses a hypercubic box with Dirichlet boundary conditions in one direction as an IR regulator, and the squared field strength after a certain depth of gradient flow
\cite{Narayanan:2006rf,Luscher:2010iy}
as the observable used to match the two lattices.
The precise matching for fermions is not specified in Ref.~\cite{Nada:2020jay},
because the reference only considers pure glue QCD.

For applications to the UV behavior of the gauge coupling, the fermion mass renormalization is not particularly important, since all quarks are light compared to the lattice spacing scale.
But for our application the quark mass matching is important.
We will make three changes to the procedure of \cite{Nada:2020jay}:
\begin{itemize}
    \item We will use boxes with a large extent along one direction, to allow the comparison of physical quark masses in terms of a meson mass plateau.
    \item We will attempt to use information from a range of gradient flow depths, rather than a single gradient flow depth, to simultaneously achieve high statistics and to measure and extrapolate away lattice-spacing effects.
    \item We will use thermal, rather than Dirichlet, boundary conditions in one direction.
    This restores lattice translation invariance, which improves statistics.
    But the main reason has to do with our familiarity with thermal boundary conditions, and we feel that Dirichlet boundaries would also be a good choice.
\end{itemize}

In the next section we will lay out our procedure for step-scaling with finite-temperature boundary conditions in the simplified case of pure-glue QCD, which will allow us to test it in a context where accurate results are already known.
Then in Section \ref{sec:milc}, we will present our analysis of scale setting with 2+1+1 flavors of HISQ fermions.
We end with some brief conclusions and outlook.

\section{Methodology and Pure-Glue Test \label{sec:gf}}

The QCD coupling varies with scale, shrinking in the UV and growing large in the IR \cite{Gross:1973id,Politzer:1973fx}.
A particularly clean way to measure this is to consider the gauge fields after the application of gradient flow.
Gradient flow is an operation on the gauge fields defined through a gauge-covariant diffusion equation
\cite{Luscher:2010iy,Luscher:2011bx}:
\begin{align}
\label{eq:contgr}
  \partial_{\tauf} B_{\nu}(x,\tauf) = D_{\mu} G_{\mu \nu}, \quad B_{\mu}(x,0)|_{\tauf=0} = A_{\mu}.
\end{align}
The covariant derivative $D_{\mu}$ and the field strength tensor $G_{\mu \nu}$ are defined as:
\begin{eqnarray}
\label{GandD}
   G_{\mu \nu} &=& \partial_{\mu} B_{\nu} -  \partial_{\nu} B_{\mu} + [B_{\mu},B_{\nu}] \\ \nonumber
   D_{\mu} X &=& \partial_{\mu} X + [B_{\mu},X] .
\end{eqnarray}
The partial derivative $\partial_{\tauf}$ is with respect to a dimensionful parameter $\tauf$ which is understood as the flow time/depth.
At leading perturbative order, gradient flow is equivalent to convolution with a Gau{\ss}ian \cite{Luscher:2010iy}
\begin{align}
B_{\mu}(x,\tauf) = \int d^{4}y K(x-y) A_{\mu}(y), \quad K(z) = \frac{e^{-z^{2}/4\tauf}}{(4\pi \tauf)^{2}}
\end{align}
or, in momentum space,
\begin{equation}
    B_\mu(p,\tauf) = A_\mu(p) e^{-p^2 \tauf} \,.
\end{equation}
The expectation value of the squared field strength
$E_{t} \equiv \Tr G_{\mu \nu} G_{\mu \nu}/2$
is then dominated by the most UV scale where fields retain their fluctuations, $p \sim 1/\sqrt{\tauf}$, and a leading-order calculation shows that
\cite{Luscher:2010iy}
\begin{align}\label{eq:gfcont}
\langle E_{t} \rangle = \frac{3(N_c^2-1)g^2}{2} \int \frac{d^4 p}{(2\pi)^4} e^{-2\tauf p^2}
= \frac{24}{128 \pi^{2}} \frac{g^{2}}{\tauf^{2}} \,.
\end{align}
This expression can be used to \textsl{define} the gauge coupling at the scale $\mu = 1/\sqrt{8\tauf}$ in the gradient-flow scheme, 
\begin{equation}
\label{defgflow}
    \gflow(\tauf) \equiv \frac{128\pi^2 \tauf^2 \langle E_t \rangle}{24} \,.
\end{equation}
The relation between this coupling and the \MSbar\ coupling $g^2(\mubar)$ is known to NNLO \cite{Harlander:2016vzb}, and it indicates that $\gflow(\tauf)$ is quite close to $g^2(\mubar = 1/\sqrt{8\tauf})$.
Evaluating $\langle E_t\rangle$ can therefore be thought of as a way of mapping out the scale dependence of the gauge coupling, with a specific value of $\gflow$\ determining a unique \textsl{physical} scale $1/\sqrt{8\tauf}$, in vacuum and in the continuum.

\begin{figure}[htb]
\centering
\includegraphics[width=0.48\textwidth]{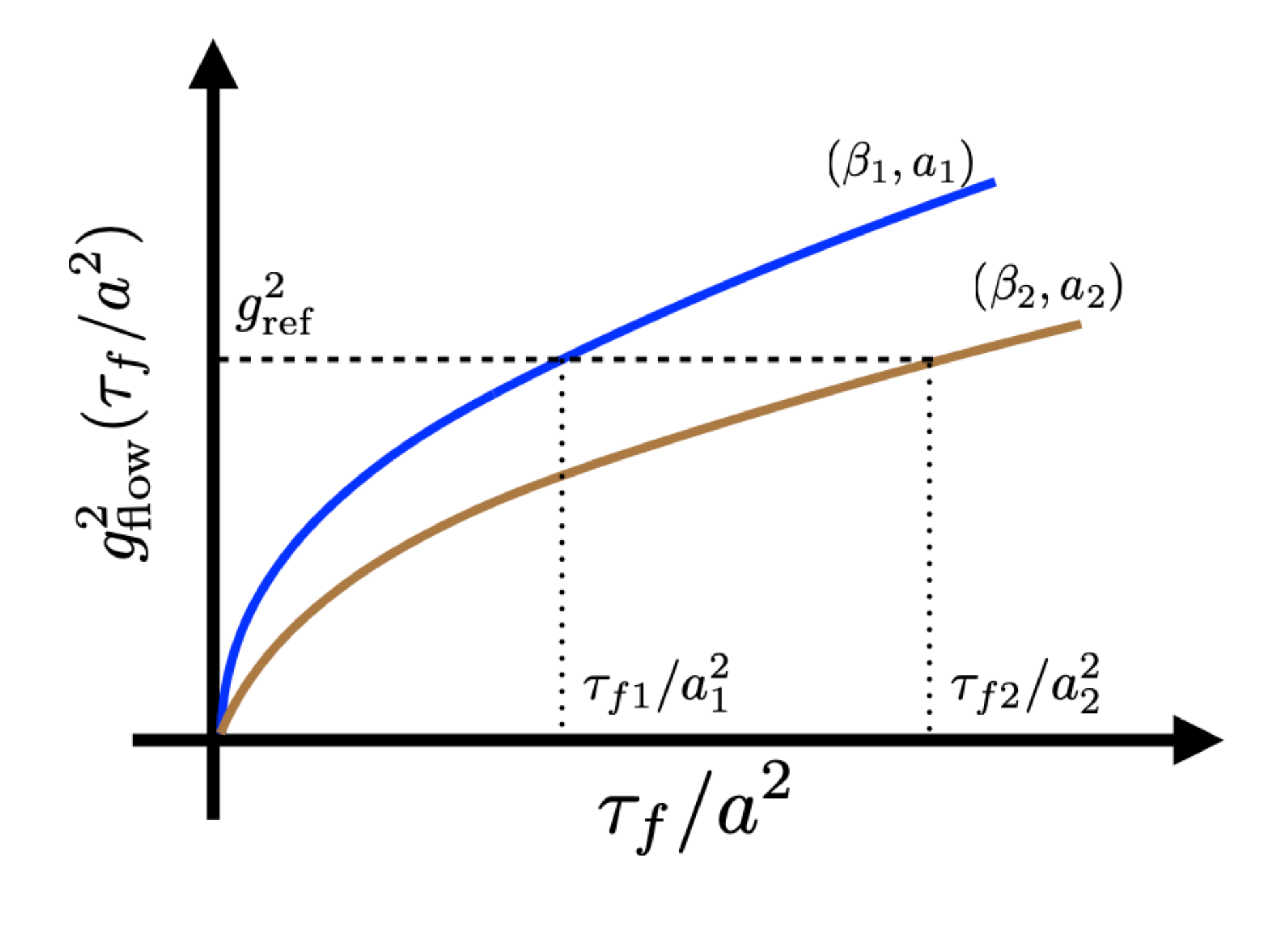}
\caption{\label{fig:cart} Cartoon showing the matching of coupling $\gflow(\tauf/a^2)$ for two lattices  of bare coupling and spacing ($\beta_1,a_1$) and ($\beta_2,a_2$).
A given $\gflow$ value is crossed at different flow depths as measured in lattice units.
These should coincide with the same physical flow depth.
Therefore, the rescaling of the horizontal axis needed to make the curves overlap determines the squared lattice spacing ratio.}
\end{figure}

To match the scales between two lattices, one can in principle choose a specific value of $\gflow$, find the flow depth $\tauf$ in lattice units which it takes to achieve this value, and equate these two $\tauf$ values.
This is the philosophy used, for instance, in Ref.~\cite{Luscher:2010iy,BMW:2012hcm}, and we illustrate it in Figure \ref{fig:cart}.
There are two potential issues with this approach.
First, lattice artifacts contaminate the results on different lattices by different amounts.
Second, the approach requires a volume large enough to see large-volume vacuum behavior.
The former problem can be handled approximately by choosing a combination of action, gradient flow methodology, and lattice observable which minimizes $a^2$ and even $a^4$ errors -- though it is difficult to do so beyond leading perturbative order.
Therefore, we will show how to use the result over a range of flow depths to capture and eliminate such spacing errors.
The latter problem can be solved by introducing a finite box size as an IR regulator and ensuring that precisely the same box size is used for each lattice.
The relation between $\tauf^2 \langle E_t \rangle$ and scale is then dependent on this box size and does not represent a universal curve in QCD, but it can nevertheless be computed at one known lattice spacing and used to relate to another lattice spacing.
This approach is nearly universal in step-scaling calculations, which frequently use Dirichlet boundary conditions in one direction.
We will use thermal boundary conditions instead.
We will now describe how we handle each problem in a little more detail.

\subsection{Lattice spacing effects}

A detailed study of tree-level lattice spacing corrections was performed in Ref.~\cite{Fodor:2014cpa}.
The paper studied effects of three choices on $\langle E_t \rangle$, namely,
the choice of the lattice action used to generate the ensemble,
the choice of action and procedure in carrying out the gradient flow,
and the choice of observable used to measure $E_t$.
At leading order the effect of the fermion implementation does not enter into a bosonic variable such as $E_t$ and is therefore not considered.
Therefore one expects lattice-spacing effects to appear at even powers of the lattice spacing:
\begin{align}
\la E_{t} \ra_{\mathrm{latt}} = &\la E_{t} \ra_{\text{cont}} \left( 1 +\frac{a^{2}}{\tauf} C_{1} + \frac{a^{4}}{\tauf^{2}} C_{2} +  \mathcal{O}\left(\frac{a^{6}}{\tauf^3}\right) \right) 
\nonumber \\ 
\Rlatt \equiv & \frac{\la E_t \ra_{\mathrm{latt}}}
{\la E_t \ra_{\mathrm{cont}}}
 = \left( 1 + \frac{a^2}{\tauf} C_1 + \frac{a^4}{\tauf^2} C_2 + \ldots \right)\,.
 \label{Rlatt}
\end{align}
\begin{table}[t]
\renewcommand{\arraystretch}{1.7}
\begin{centering}
\begin{tabular}{c|c|c}
\hline \hline
Action-flow-observable & $C_{1}$ & $C_{2}$ \\
\hline
W-W-W & $\frac{1}{8}$ & $\frac{3}{128}$ \\
\hline
W-W-C & $- \frac{1}{24}$ & $ - \frac{1}{512}$ \\
\hline
W-Z-W & $0$ & $ \frac{101}{3840}$ \\
\hline
W-W-LW & $\frac{13}{72}$ & $ \frac{13}{384}$ \\
\hline
W-Z-LW & $\frac{1}{18}$ & $ \frac{17}{768}$ \\
\hline
W-Z-iC & $\frac{1}{18}$ & $- \frac{7}{512}$ \\
\hline
LW-Z-iC & 0 & $-\frac{3}{120}$ \\
\hline
LW-Z-LW& 0 & $\frac{101}{3840}$ \\
\hline
\end{tabular}
\caption{\label{tab:gfimp} The improvement coefficients $C_{1}$ and $C_{2}$
in terms of the generating action, the flow procedure, and the observable.
The label W means Wilson, C is Clover, Z is Zeuthen flow, iC is improved Clover and LW is L\"uscher Weisz.}
\end{centering}
\end{table}
Here we introduce $\Rlatt$ as the ratio, which describes lattice-continuum effects.

Improving the action and observable follows well-known procedures
\cite{Luscher:1984xn}.
Improving the gradient flow procedure requires slightly more than using an improved definition of $G_{\mu\nu}$ in \Eq{GandD}, as recently explained in
Ref.~\cite{Ramos:2015baa}.  The full procedure is called \textsl{Zeuthen flow} and we have implemented it in addition to Wilson flow for our analysis.
This subtlety was discovered after the work of Ref.~\cite{Fodor:2014cpa}.
But recently the $C_{1}$ and $C_{2}$ coefficients for different combinations of action, flow, and operator choices, including Zeuthen flow, were computed in \cite{tuprints23185} and are listed in the Table.~\ref{tab:gfimp}.

In analyzing pure-glue QCD we will use the Wilson action, and will therefore be most interested in the W-Z-W combination,
but HISQ fermions are paired with L\"uscher-Weisz-type improved gauge actions and therefore our analysis of full QCD will use the last two combinations in the table.

\subsection{Thermal corrections}

Introducing an IR regulator, such as a small box with nontrivial boundary conditions, changes $\gflow(\tauf)$ from its vacuum value.
Provided that all simulations are performed with precisely the same IR regulator, this does not matter; we can still match different lattice spacings to establish their ratio.
However, we don't know the lattice spacing ratio \textsl{a priori} -- in fact, this is the whole point -- so until we converge to the correct parameters, the different lattices we study will generically not have precisely the same IR regulation.
For instance, if two lattices are supposed to have a factor-of-2 lattice spacing ratio, we might study one on an $8\times 32^3$ box and the other on a $16\times 64^3$ box.  But if the true ratio turns out to be 1.89 rather than 2.00, then these will represent different physical temperatures and therefore different expected $\gflow(\tauf)$ values at large $\tauf$.
Therefore it is good to determine everything we can about the box-size effects on $\langle E_t \rangle$, to take them into account in the matching, and to stick to a range of $\tauf$ values where the effects are not large.

With this in mind, we will calculate the thermal corrections to $\langle E_t \rangle$.
We will do so first at lowest perturbative order, where the corrections prove to be \textsl{exponentially} small for small $\tauf T^2$.
But at higher orders in the coupling there are \textsl{polynomial} in $\tauf T^2$ effects, which we will also be able to determine.
Since we are interested in infrared effects here, we will work in the continuum.

\begin{figure}[H]
\centering
  \includegraphics[scale=0.53]{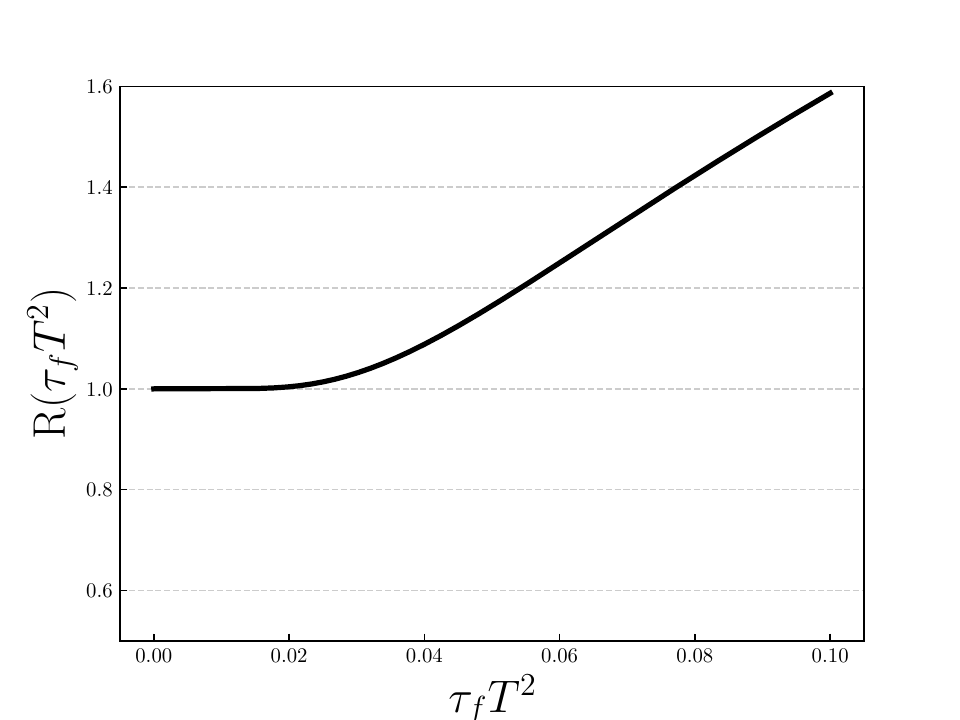}
  \caption{\label{fig:therm} Thermal corrections $R_{T,\mathrm{LO}}(\tauf T^2)$ from \Eq{R_LO}}
\end{figure}

At the free theory level, the expectation value of the field strength with thermal boundary conditions is:
\begin{align}
\langle E_{t} \rangle(T) = 12 g^{2} T \sum_{p_{0}=2 \pi n T} \int \frac{d^{3}p}{(2\pi)^{3}} e^{-2\tauf (p^{2}_{0} + p^{2})} .
\end{align}
This can be evaluated using the Poisson summation formula, leading to%
\footnote{This result does not directly appear in the reference, but can be found by combining the reference's Eqs.(3.3,3.4).}
\cite{Eller:2018yje}
\begin{align}\label{eq:therm}
\langle E_{t} \rangle (T) = \langle E_{t} \rangle (T=0) \sum_{n \in Z} e^{-n^2 / 8 \tauf T^2} \,.
\end{align}
Provided that $8\tauf T^2 \ll 1$, this result is \textsl{exponentially} close to 1.
We name the ratio of the thermal to the vacuum squared field strength
\begin{equation}
\label{R_LO}
    R_T \equiv \frac{\langle E_t\rangle (T)}{\langle E_t \rangle(T=0)} \,, \quad
    R_{T,\mathrm{LO}} = 1 + 2 \sum_{n=1}^{\infty} e^{-n^2/8\tauf T^2} .
\end{equation}
This result is shown in Figure \ref{fig:therm}.
If a spatial extent is comparably small to the time extent, the full correction is a product of this correction times a factor for each space direction, with the length $L$ playing the role of $1/T$.

We see that the thermal effects are negligible up to $\tauf T^2 \sim 0.02$ and become important after $\tauf T^2 \sim 0.04$ hitting the 20\% level at $\tauf T^2 \sim 0.05$. 
Therefore we will typically avoid flow values larger than $\tauf T^2 = 0.03$, so that these corrections remain small.

Besides these effects, there are effects at higher order in the coupling which are only polynomially suppressed in the temperature,
$R \sim g^2 (\tauf T^2)^2$.
To see this, consider the relation between $\langle E \rangle$ and the trace anomaly:
\begin{align} 
\label{eq:freeE}
	\epsilon - 3P &= \frac{T}{V} \frac{d \; \mathrm{ln} Z}{d \; \text{ln} \mu } \\ \nonumber
	&= \frac{1}{g^4} \frac{dg^2}{d \mathrm{ln} \mu} \langle E \rangle .
\end{align}
The thermal contributions to the pressure have a known perturbative expansion at high temperatures,
\begin{equation}
    P(T) = (A + B g^2(\mu=\pi T) + \ldots) T^4,
\end{equation}
where the coefficients $(A,B)$ and higher order terms can be found, for instance, in Refs.~\cite{Arnold:1994ps,Braaten:1995jr,Kajantie:2002wa}.
The trace of the stress tensor is related to the pressure via
\begin{equation}
    \epsilon-3P = T \frac{dP}{dT} - 4 P =
 BT^4  \frac{\partial g^2}{\partial \ln \mu}.
\end{equation}
Combining with \Eq{eq:freeE}, we find
\begin{align}
\langle E_t \rangle(T) = {}&{} \langle E_t \rangle(T=0)
+ g^4 B T^4 \,,
\nonumber \\
R_T = {}&{} R_{T,\mathrm{LO}} + \frac{8\pi^2 g^2}{9} (\tauf T^2)^2 \;
\mbox{(for pure glue)}.
\label{fullRT}
\end{align}
Here we used the standard pure-glue expression for B from \cite{Arnold:1994ps}.
In the presence of light fermions the correction is multiplied by $1+5n_f/12$ with $n_f$ the number of species with $m<\pi T$ \cite{Arnold:1994eb}.
We also assume that the thermal-to-vacuum difference in $\langle E \rangle$ is not influenced by gradient flow, which will work until the leading-order corrections we discussed above begin to play a role.

Combining these corrections, we account for lattice-spacing and thermal effects by using
\begin{equation}
\label{gflowcorr}
    \gflow(\tauf) = \frac{128 \pi^2}{24\tauf^2} 
    \frac{\langle E_t \rangle_{T,\mathrm{latt}}}{\Rlatt R_T} + \mbox{(higher order)} .
\end{equation}
In practice we apply the two corrections as additive shifts, that is, we use $1+(\Rlatt{-}1) + (R_T{-}1)$ rather than $\Rlatt R_T$.

\subsection{Application to pure-glue QCD}

Our procedure for scale setting is then the following:
\begin{enumerate}
    \item 
    Choose a reference lattice-spacing which is already known.
    \item 
    Choose a box size which will be an integer number of lattice spacings in each direction.
    \item
    Choose a second desired lattice spacing such that the same box will also be an integer number of spacings in each direction.
    \item
    Make a first estimate (based on extrapolating known results) for the coupling $\beta$ to use with this lattice.
    \item 
    Measure $\gflow$, including spacing and temperature corrections, for each spacing in this box.
    \item
    Choose a range of $\tauf$ values where each $\gflow$ result should be relatively free of lattice-spacing and temperature effects.
    \item
    Determine the flow-rescaling which causes the $\gflow$ results to most accurately overlap in our $\tauf$ window, and therefore determines the true ratio of lattice spacings.
    \item 
    If the lattice spacing ratio is too far from the desired value, the volumes are inequivalent.  
    In this case we make an improved $\beta$ estimate and repeat the procedure.
\end{enumerate}
Before applying this procedure in 2+1+1 flavor QCD, we will first test it in the pure-glue case, where it is not necessary to match quark masses and where precise scale setting has already been established.
Since the most accurate scale setting has been performed for the Wilson action
\cite{Francis:2015lha,Burnier:2017bod},
we will also use this action choice.

To test our procedure, we choose a reference lattice with lattice coupling $\beta_1$, corresponding to a spacing $a_1$, such that a lattice with temporal extent $N_\tau = 10$ will correspond to a temperature of $T=4.1\,T_c$.
We then try to identify what lattice couplings $\beta_2,\beta_3$ will be at the same temperature on lattices of temporal extent $N_{\tau} = 12,14$.
We adopt as our ``guess'' for the correct $\beta_2,\beta_3$ the values determined by the pocket formula provided in Ref.~\cite{Burnier:2017bod}.
We then choose lattices with aspect ratios larger than 3, such that finite-volume effects are expected to be small.
The specific parameters are shown in 
Table~\ref{tab:pg}.

\begin{table}[htb]
	\renewcommand{\arraystretch}{1.2}
\begin{tabular}{c|c|c|c|c|c}
\hline \hline 
Lattice & $\beta$ & $a$ (fm) & $N_\tau$ & $N_x \times N_y \times N_z$ & $N_{\mathrm{trajs}} $\\
		\hline \hline
	$B1$ &	7.30916  & 0.01675(67) & 10 & $36 \times 32 \times 32$ & 6400 \\
		\hline 
	$B2$ &	7.46275 &  0.01396(56)& 12 & $40 \times 32 \times 36$ & 6400 \\
		\hline
	 $B3$ &	7.59354 &  0.01196(48)& 14 & $48 \times 48 \times 48$ & 6400 \\
		\hline \hline
	\end{tabular}
	\caption{\label{tab:pg} Input lattice parameters at temperature  $T=4.1 \ T_c$ in pure glue lattice QCD.}
	\end{table}

\begin{figure*}[t]
\centering
\includegraphics[scale=0.5]{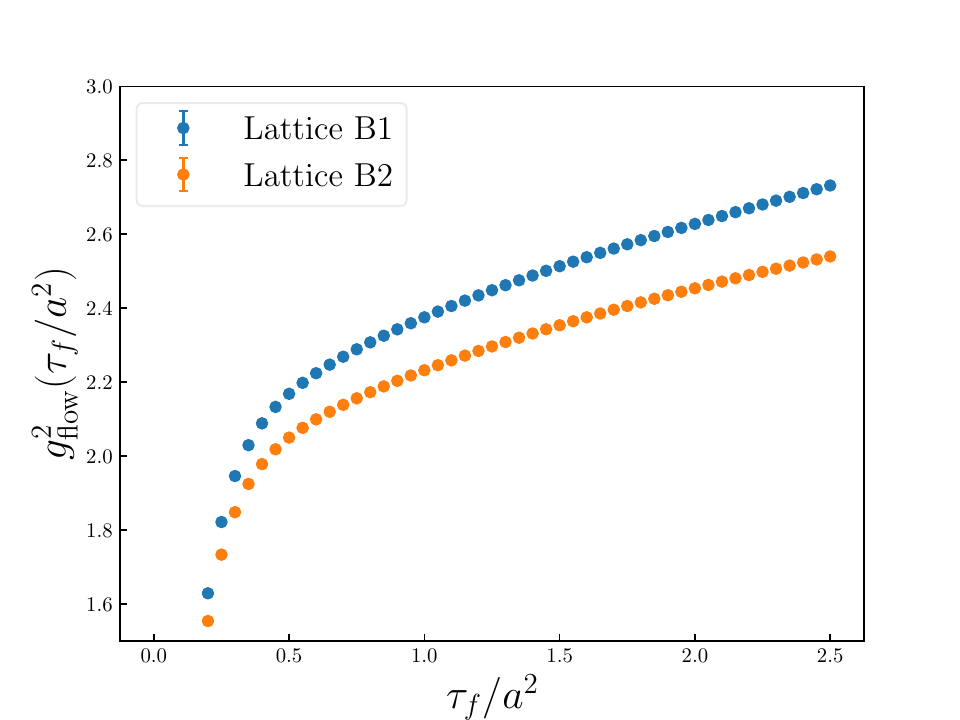}
\includegraphics[scale=0.5]{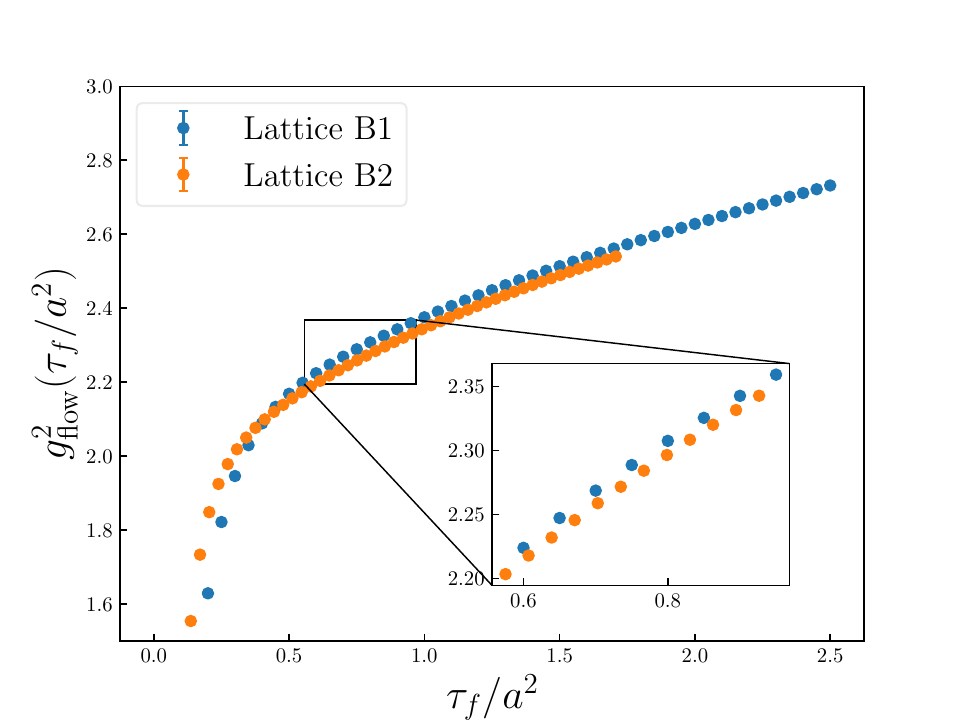}
\caption{\label{fig:pg_wflow}
  Coupling after gradient flow, defined in \Eq{defgflow} without lattice spacing or thermal corrections, for the $B1,B2$ lattices.
  Left:  expressing $\tauf$ in units of the lattice spacing.
  Right:  rescaling $\tauf$ for the $B2$ lattice (and therefore the $x$-axis) to optimize the overlap of the two curves.  }
\end{figure*}

We then evaluate $\gflow$ at a range of gradient flow depths for each lattice.
The result for lattices $B_1,B_2$ is shown in Figure \ref{fig:pg_wflow}. 
The figure shows $\gflow$ from \Eq{defgflow}, \textsl{without} the lattice-spacing and thermal corrections described in \Eq{gflowcorr}, for two lattices.
The lattice $B2$ clearly has a smaller coupling value, and the scale dependence of the coupling is also manifest.
The turn-off towards zero at very small $\tauf$ represents strong lattice effects; for small $\tauf$ the integral in \Eq{eq:gfcont} is dominated by $p$-values which are absent on the lattice since the lattice $p$ is restricted to a Brillouin zone.
Ignoring this region, 
a rescaling of $\tauf$ for $B2$ relative to $B1$ brings the two curves into approximate agreement.
Specifically, if we compare $\gflow$ for $B1$ at flow depth $\taufa = K a_1^2$
with $\gflow$ for $B_2$ at flow depth $\taufb = s^2 K a_2^2$ we find them to agree, for a range of $K$ at a fixed $s$ value.
This implies that the physical scales are the same,
$a_1^2 = s^2 a_2^2$ or $s = a_1/a_2$.

The complication, also illustrated in Figure \ref{fig:pg_wflow}, is that the curves never agree completely over a wide range.
At small $\tauf$ the lattice spacing effects are important, while at large $\tauf$ the error bars are larger and thermal effects come into play.
What we want is the $s$ value which optimally rescales the flow depths such that the agreement is optimized over a range of $\tauf$ values.
We now outline our procedure for doing so.

We generate a set of gauge field configurations for each lattice $B1,B2,B3$.
For each generated configuration we measure $\gflow$ at a discrete series of flow depths $\taufi$.
These are corrected for spacing and temperature effects using \Eq{gflowcorr}.
We will fit using the data within a flow range $\tauf > 1.0 a^2$ and $\tauf < 0.04/T^2$, both for the coarser (smaller-$N_\tau$) lattice.
While we determine the statistical errors for each $\gflow(\tauf)$ value, these will be highly correlated between nearby $\tauf$ values, so the errors are always determined by breaking the Markov-chain data streams into blocks, larger than the autocorrelation time at the largest $\tauf$ we use, and applying the jackknife method to the full fitting procedure.

One complication is that
we want to compare $\gflow(B1,\taufa)$ to
$\gflow(B2,\taufb=s^2 \taufa)$ with $s$ a real-valued fitting parameter.  But we have only determined $\gflow(\tauf)$ at discrete values, and for generic $s$ values these will not ``line up'' between the two data sets.
To handle this, rather than comparing $\gflow(B1)$ to $\gflow(B2)$ directly, we will compare \textsl{each} to a model for $\gflow$.
Specifically, we define $\gfit(\tauf)$ to be a few-parameter piecewise-cubic Hermite spline function.
Define $s_0$ to be the expected value of the lattice-spacing ratio.
The optimal $s$ value is the one which best allows $\gfit$ to match $\gflow(\taufa) = \gfit(\tauf=\taufa)$ 
\textsl{and} to fit $\gflow(\taufb) = \gfit(\tauf = \taufb/s^2)$.
Therefore, in comparing, say, lattice $B1$ with lattice $B2$, we define a chisquare value which is
\begin{align}
    \chi^2 = & \sum_{\tau_i \in [\tau_{\mathrm{min}},\tau_{\mathrm{max}}]}
    \frac{\Delta \tau_i}{\tau_i} 
    \frac{(\gflow(B1,\tau_i) - \gfit(\tau_i))^2}
    {\sigma^2_{B1,\tau_i}}
    \nonumber \\
    & {} + \!\!\sum_{\tau_i \in [ s_0^2 \tau_{\mathrm{min}}, s_0^2 \tau_{\mathrm{max}}]}
    \frac{\Delta \tau_i}{\tau_i}
    \frac{(\gflow(B2,\tau_i) - \gfit(\tau_i/s^2))^2}
    {\sigma^2_{B2,\tau_i}} \,,
\end{align}
that is, the chisquare compares the deviation of $\gflow$ measured on $B1$ to a spline fit \textsl{and} the deviation of $\gflow$ measured on $B2$ to the same spline after a $\tau$-rescaling.
The fit is optimized over both the rescaling $s$ and the parameters of the spline fit.
The factor $\Delta \tau/\tau$, with $\Delta \tau$ the spacing between the measured $\tau$ values, accounts for the high degree of autocorrelation between $\gflow$ at nearby flow depths.
In addition, rather than using the coefficients $C_1,C_2$ of \Eq{Rlatt} as given in the table, we instead include $C_1$ as a fitting parameter with a weak prior favoring the leading-order value, and neglect $C_2$.

We have carried out this procedure twice, with two choices of observable.
We always use the Wilson action and the improved Zeuthen flow.
But we apply the procedure once using the Wilson observable and once using the improved-clover observable.
Table \ref{tab:gfimp} shows that the former case has $C_1=0$, while in the latter case the corrections from the Wilson action are uncorrected and $C_1 = 1/18$.
The results using each action choice are listed in Table ~\ref{tab:pg}.
The fits are performed on all pairs of generated lattices from Table~\ref{tab:pg}.
The relevant scale ratio of the pairs are already known from Ref.~\cite{Francis:2015lha} and are denoted by the input scale ratio $s_{\mathrm{input}}$.
\begin{table}[ht]
	\renewcommand{\arraystretch}{1.2}
\begin{tabular}{c|c|c|c|c|c}
\hline \hline 
Lattice 1 & Lattice 2 & $s_{\mathrm{obs}}$ & $s_{\mathrm{input}}$  & $C_1$ & Obs \\ 
\hline\hline
  $B1$ & $B2$ & 1.208(8) & 1.1998(9) & 0.0498(17)  & iC \\
  \hline
  $B2$ & $B1$ & 1.210(5) & 1.1998(9) & 0.0498(17)  & iC \\
  \hline
  $B1$ & $B3$ & 1.414(10) & 1.4005(8) & 0.0326(31)  & iC \\
  \hline
  $B2$ & $B3$ & 1.167(7) & 1.1672(7) & 0.0526(36)  & iC \\
  \hline
  \hline
  $B1$ & $B2$ & 1.249(9) & 1.1998(9) & 0.0019(18) &  W \\
  \hline
  $B1$ & $B3$ & 1.465(9) & 1.4005(8) & 0.0041(6) &  W \\
  \hline
  $B2$ & $B3$ & 1.187(5) & 1.1672(7) & 0.0069(11) &  W \\
  \hline \hline 
\end{tabular}
\caption{\label{tab:pg_results} Fit results for the lattices $B1$, $B2$ and $B3$ with Zeuthen flow and the observables  iC (improved Clover), W (Wilson).}
\end{table}
In our fitting procedure, the spline is initialized from the data on one or the other of the two lattices.
Therefore as a first check of the procedure, we establish the consistency of  fits under the exchange of the two ensembles in a pair.
This is shown in the first two lines of Table \ref{tab:pg_results} with ensembles B1 and B2.
We see that the procedure is robust against this difference.

The determined value of the parameter $C_1$ is surprisingly close to its leading-order estimate.
This suggests that tailoring the action-flow-observable combination to have small $C_1,C_2$ is a viable strategy.
We will use this approach in the next section.
The table also indicates that the combination Wilson-Zeuthen-Wilson gives larger final deviations in the determined scale ratio.
Based on examining the final fitted curves, we believe that this is because of the larger $a^4$ corrections which occur in this case.
With our choice of $\tauf > 1.0 a^2$, such corrections are not yet negligible at the lowest edge of our flow range, and our failure to account for them systematically biases our results.
This implies that it may be necessary to try to tune away $a^4$ effects, and/or to account for them in the fit.
We will do so in the next section.

\section{Extension of MILC scale setting \label{sec:milc}}

\begin{table}[hbt]
	\centering
	\begin{tabular}{|c|c|c|c|}
		\hline
		$\beta$ &$a$(fm) & $am_c$\\ 
		\hline \hline
		6.72 & 0.05662(13) & 0.2679(1)  \\ \hline
		 7.28 & 0.03215(13) & 0.1333(1) \\ \hline
	\end{tabular}
	\caption{\label{tab:reflats}Lattice parameters for two lattices, as determined by the MILC collaboration \cite{Bazavov:2017lyh}.
 Note that the $\beta=7.28$ case was calculated at an unphysically heavy pion mass.}
\end{table}

Our eventual goal, as explained in the introduction, is to study the topological susceptibility of QCD in the temperature range 400--1100 MeV.
To do so we will choose three target temperatures and we will design lattices ($\beta$ and $m_q$ combinations) at the physical point and at a set of lattice spacings corresponding to $N_\tau = 8,10,12,14$ at each desired temperature.
The two finest 2+1+1 flavor MILC lattices to be precisely scale-set
\cite{MILC:2012znn,Bazavov:2017lyh} are shown in Table \ref{tab:reflats}.
The scale ratio of these two lattices happens to be $a_1/a_2 \simeq 8/14$.
Therefore we can choose a temperature $T=435\,\mathrm{MeV}$ for which these known lattices correspond to the $N_\tau = 8$ and $N_\tau=14$ lattices; we only need to determine the $N_\tau = 10,12$ lattices which lie between them.
Going to higher temperatures, we will also investigate
$T = 682\,\mathrm{MeV}$, for which the finer MILC lattice has $N_\tau = 9$,
and $T = 1022\,\mathrm{MeV}$, for which the finer MILC lattice has $N_\tau = 6$.
This provides three temperatures which span the range over which the topological susceptibility plays an important role in determining the axion production efficiency \cite{Klaer:2017qhr,Moore:2017ond}.
However, $N_\tau=9$ cannot be simulated with staggered quarks and $N_\tau = 6$ is too small to provide reliable lattice results; therefore in establishing the lattice-scale matching for these two choices, we will investigate temperatures which are half as hot as the target temperature, corresponding to the use of double the $N_\tau$ value which we intend to employ in the eventual topology simulations.
This will also be important when we match $m_c$ values, as we need $m_c \gtrsim \pi T$.
For this reason, we will target the scale determination of the lattices listed in Table \ref{tab:scalelatts}.

\begin{table}[ht]
\centering
\caption{\label{tab:scalelatts} Lattice parameters for the scale setting calculation.
 Values in bold are known from [4,5]; Other parameters are final results of the calculation.}
\begin{tabular}{|c|c|c|c|c|c|}
 \hline
T(MeV) & Lattice & $\beta_{\mathrm{tuned}}$ & $a m_c$ & $\epsilon$ & $\tau$\\ \hline
\multirow{4}{*}{435(1)} &$\mathbf{8 \times 16^2 \times 64 } $  & \textbf{6.720}  & 0.2679(1) & 0.045 & 1.47\\
 & $10 \times 20^2 \times 72$ &  6.950 & 0.1994(13)  & 0.04 & 1.20\\
 & $12 \times 24^2 \times 96$ &  7.130 & 0.1636(13) & 0.04 & 1.20 \\ 
		
\hline
 \multirow{5}{*}{341(2)}   & $\mathbf{18 \times 18^2 \times 120}$  &  \textbf{7.280}  & 0.1333(1) & 0.04 & 1.00\\
 & $16 \times 16^2 \times 96$  &  7.150   & 0.1432(30) & 0.06 & 1.08  \\
 & $20 \times 20^2\times 120$  &  7.390   & 0.1107(25) & 0.05 & 1.11 \\
 & $24 \times 24^2 \times 144$  &  \textit{7.600} & 0.0884(45)& -& -  \\ 
 & $28 \times 28^2 \times 168$  &  7.715  & 0.0827(17) & 0.025 & 1.25\\ 
 \hline
 \multirow{5}{*}{511(2)}   & $\mathbf{12 \times 12^2 \times 72}$  &  \textbf{7.280} & 0.1333(1)  & 0.04 & 1.20 \\
 & $16 \times 16^2 \times 96$  &  7.600  & 0.0884(45) & 0.04 & 1.20 \\
 & $20 \times 20^2\times 120$  &  7.820  & 0.0760(35) & 0.035  & 1.47 \\
 & $24 \times 24^2 \times 144$  &  8.045    & 0.0697(46) & 0.035 & 1.47\\ 
 & $28 \times 28^2 \times 168$  &  8.220    & 0.0559(17)& 0.025 & 1.25\\ 
\hline
\end{tabular}
 \end{table}

We will proceed in two steps.
First we will determine the lattice inverse gauge coupling $\beta$ for each desired lattice using the same technique as in the previous section and an educated guess for the quark masses.
Second, we will tune the quark masses.
The mass ratios $m_s/m_c$ and $m_l/m_s$ should remain fixed at high energy scales, since the running of the quark mass depends on the difference in anomalous dimension between the $\bar\psi \psi$ operator and the $\bar\psi \slashed{D} \psi$ operator, which depends only on the operator structure which is common between quark types.
This argument should apply provided all quarks obey $m_q a \ll 1$, which is true in the lattice-spacing range we consider.
Therefore we fix the quark masses to the ratios determined by Ref.~\cite{Bazavov:2017lyh}:
\begin{equation}
\label{massratios}
	\frac{m_c}{m_s} = 11.783(1) \quad \frac{m_s}{m_l} = 27.3.
\end{equation}
We then only need to match $m_c$ values between lattices.
We will do this by measuring the $D_s$ spatial correlator mass on the reference lattice and tuning the valence $m_c$ on all other lattices to reproduce it.
Note that the $D_s$ spatial correlator mass is temperature dependent; but we always compare lattices which are at the same temperature, so this should be a common effect, provided that we use the same lattice geometry in each case.
Because the correlator mass involves measuring correlation functions over a wide range of separations, we use lattices with a large extent along one axis.
For the large-$N_\tau$ cases, numerical expediency forces us to use rather small extents in the other spatial directions; but again, as long as these are the same for all lattices, the effects should be common and should scale out.

Since we will be extending the MILC scales, we employ the same gauge and fermionic action used by them \cite{Follana:2006rc}.
We state them here for completeness.
We use the $N_f= 2+1+1$ HISQ fermionic action, where the gauge links are Fat7 smeared along with a Naik term characterizing a derivative improvement with a third nearest neighbor term.
The charm quark is implemented in the same way as the lighter quarks.
The gauge action is one-loop Symanzik  and tadpole improved using the plaquette to determine the $u_0$ factor.
The action includes planar Wilson loops with terms of type $1\times1$ and $1\times 2$ and a parallelogram-type $1\times 1 \times 1$ term.

\begin{figure*}[ht]
\centering
\includegraphics[scale=0.5]{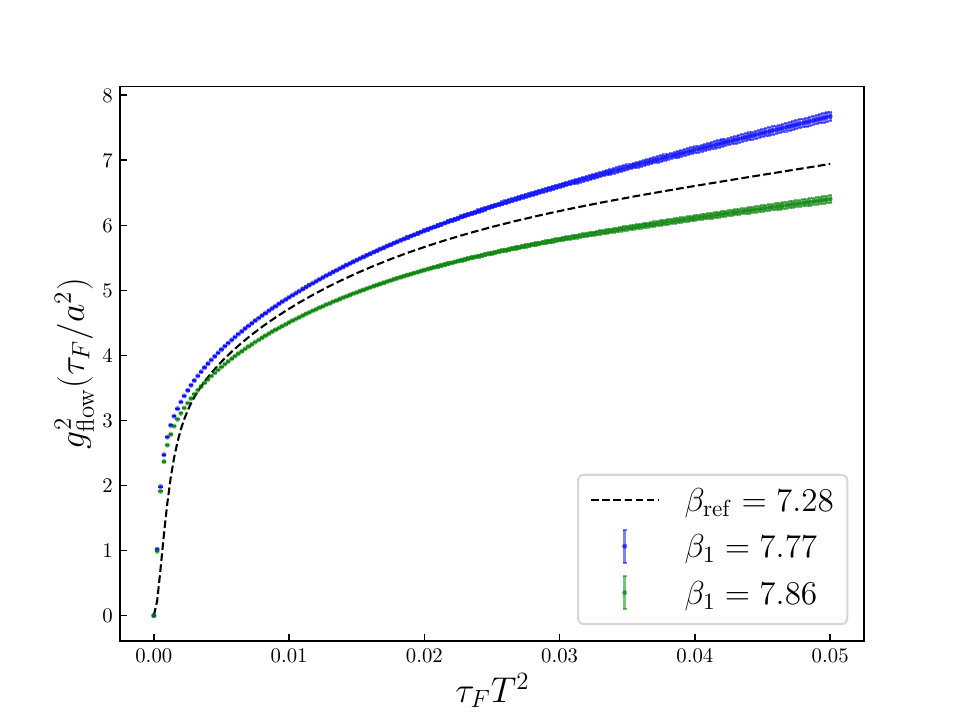}
\includegraphics[scale=0.5]{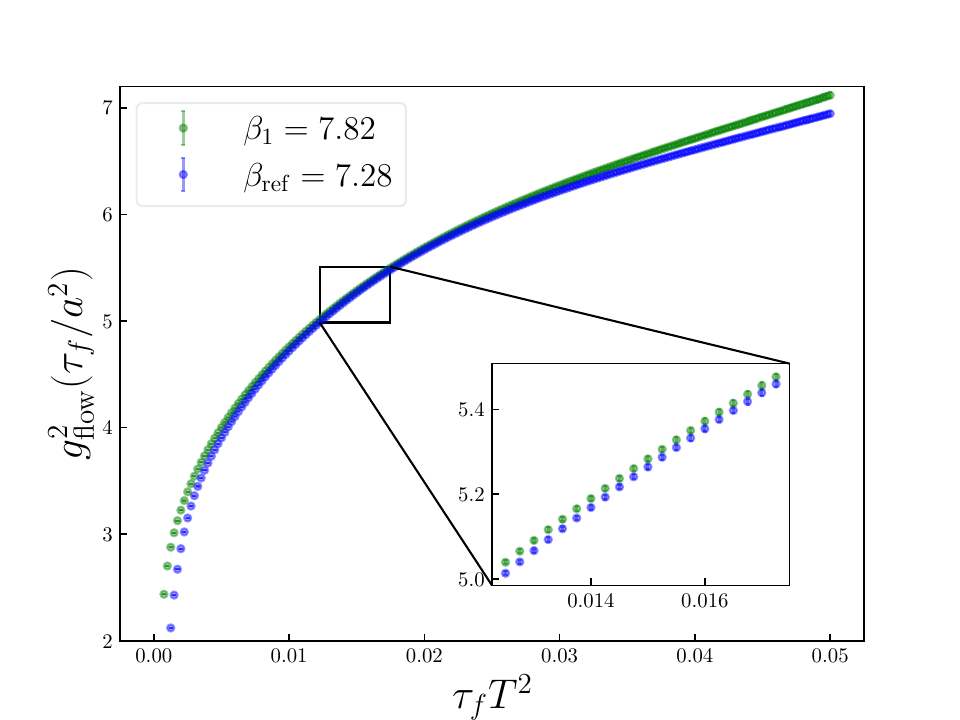}
  \caption{\label{fig:betatuning} Tuning of the gauge coupling at $T=511$ MeV, comparing lattices with $N_\tau = 12$ and $N_\tau = 20$.
  Left panel:  two $\beta$ values at $N_\tau = 20$, one too low and one too high, give incompatible results for $\gflow$ when the flow is scaled according to the target lattice spacing ratio.
  Right panel:  a $\beta$ value interpolated based on the other results gives a good match.}
\end{figure*}

\subsection{\label{sb:gsq}Tuning of gauge coupling}

In Table~\ref{tab:scalelatts}, we list the temperatures and target lattices which we will consider.
We also present the $\beta$ and $m_c$ values which emerge from our matching, and two parameters which describe the 
Rational hybrid Monte-Carlo algorithm: $\epsilon$ is the trajectory evolution step and $\tau$ is the total length of the trajectory.
The labels in bold indicate the reference parameters (listed in Table~\ref{tab:reflats}).
We always use a box length of at least $6N_\tau$ in the $z$-direction
in order to have a sufficiently long direction for a reliable computation of the screening mass in that direction, which will be relevant for the quark mass tuning.
The lattice with $\beta=7.6$, with $a = (3/4)$ times the reference value, appears for both the second and third temperature; we only use the 511 MeV temperature in establishing its parameters.

The gauge coupling tuning begins with an initial guess for the gauge coupling and charm quark mass $(\beta,m_c)$ which will correspond to the desired lattice spacing.
We note that the target lattice spacing is known since the temperature and temporal extent $N_\tau$ is fixed.
Our initial guess is based on a rational-function extrapolation of the known 2+1+1-flavor MILC scale setting values \cite{MILC:2013ops,Weber:2021hro}.
We then choose two values 1\% separated in $\beta$ which bracket this value, which we will scale set and then interpolate to find the optimal $\beta$ value.
We will then repeat the scale setting at this optimal $\beta$ value as a final check.
In each case the $m_c$ mass is varied along with $\beta$ using the rational-function relation which we establish from the prior scale setting.

We perform the scale setting itself using the same procedure as in the last section, but with the following modifications.
First, we choose a combination of action, flow, and observable such that both $C_1$ and $C_2$ vanish at tree level.
Specifically, we use the 1-loop tadpole improved gauge action which is standard for HISQ and is described above, together with improved (Zeuthen) flow.
And we use a linear combination of the L\"uscher-Weisz and improved-clover observables, chosen using the last two lines of Table \ref{tab:gfimp} such that $C_2$ cancels:
\begin{equation}
 	t^2 \langle E \rangle = \frac{ 96 t^2 \langle E \rangle_{\mathrm{LW}} 
  + 101 t^2 \langle E \rangle_{\mathrm{iC}}}{197} \,.
\label{fullimprove}
\end{equation}

We have considered three ways of treating the variables $C_1,C_2$.
In the first approach, we include them as fitting parameters, allowing them to vary independently in each matching between two lattices.
These fits reveal that both $C_1$ and $C_2$ are small, generally consistent with zero.
Also, for a given choice of lattice action, $C_1,C_2$ should be functions of the lattice coupling parameter $\beta$, which varies very little between our different lattices.
Therefore, even if nonzero, we should expect $C_1,C_2$ to be almost the same across all simulations.
So a second approach is to treat them as fitting parameters which are \textsl{common} across every fit -- thereby reducing the number of fitting parameters which must be considered.
Doing so, we find that both parameters are indeed small and consistent with zero.
This suggests a third procedure; simply assuming that the lattice improvement was successful and that their values are 0.
Somewhat surprisingly, our results both for scale ratios and for their errors are almost the same in each of these procedures.
Therefore, conservatively, we have adopted the fits in which we include $C_1,C_2$ as fitting variables.

Another change we make involves the $\tauf T^2$ range we use in the matching.
For the second and third temperatures listed in Table \ref{tab:scalelatts}, the small space extent along two axes means that the thermal correction $R_T$ calculated in \Eq{fullRT} cannot be trusted.
Rather, there is a correction of the same form as \Eq{R_LO} associated with time and with each small space extent.
We choose a limit $\tauf T^2 \leq 0.03$, where \Eq{R_LO} indicates a correction smaller than 1\%, to control these geometry-induced errors.
(In these space directions we apply periodic boundary conditions for both the gauge fields and the fermionic fields, in contrast to the antiperiodic boundaries for fermions in the time direction.)

To illustrate the way we interpolate to get to the optimal gauge coupling, consider Figure \ref{fig:betatuning}.
The figure shows a comparison of the 511 MeV lattices with $N_\tau = 12$ and $N_\tau = 20$.
On the left, we rescale based on the box size, and see that two $\beta$ values for the $N_\tau = 20$ lattice are respectively above and below the reference lattice.
By measuring the required rescaling in each case and interpolating,
we estimate that $\beta=7.820$ at $N_\tau = 20$ will be a good match for the reference lattice.
This proves correct, as seen in the right panel.
One also observes that the curves at small $\tauf T^2$ differ due to differing lattice-spacing effects.
At the largest $\tauf T^2$ values the lattices also differ, but the statistics are poorer here, and it is also beyond our cutoff $\tauf T^2 < 0.03$ where geometry effects may come into play.

\begin{figure*}[t]
\centering
\includegraphics[scale=0.43]{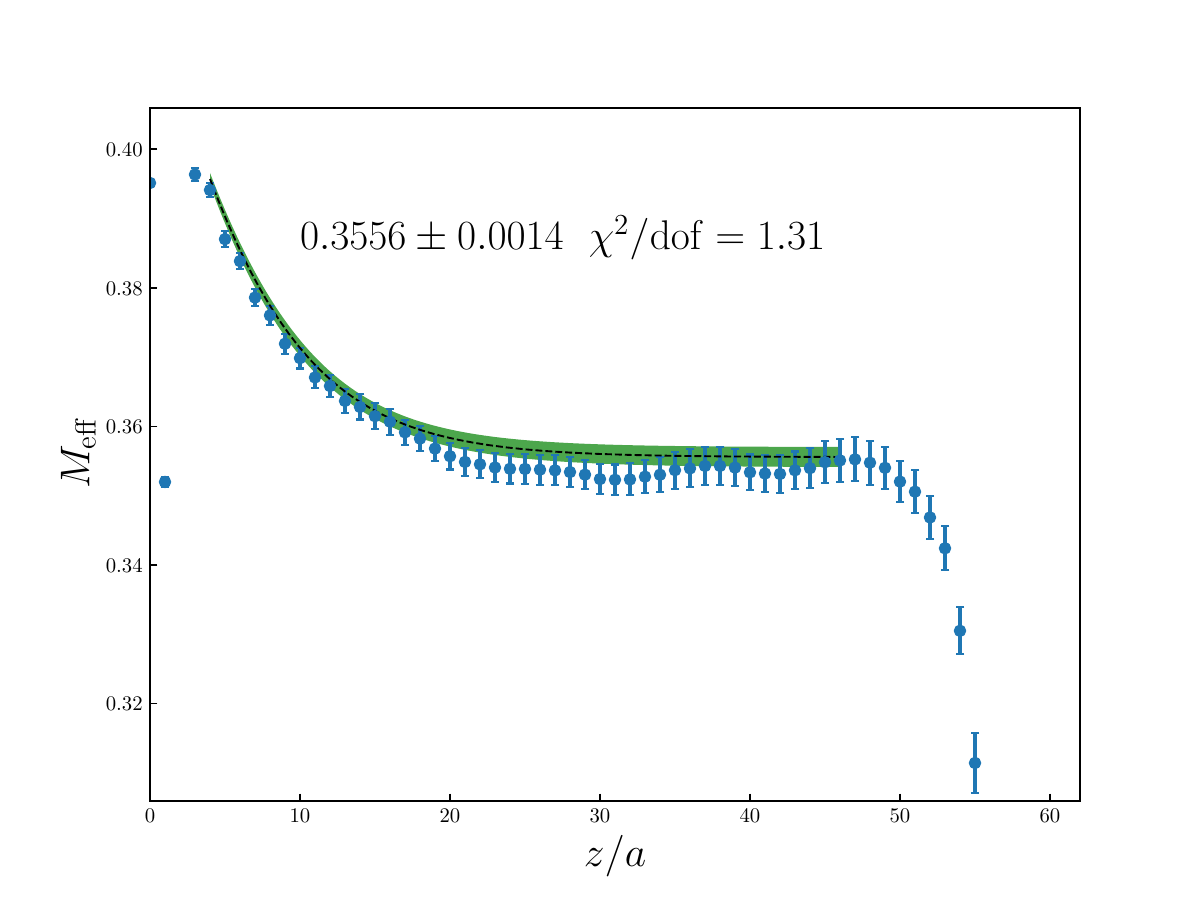}
\includegraphics[scale=0.43]{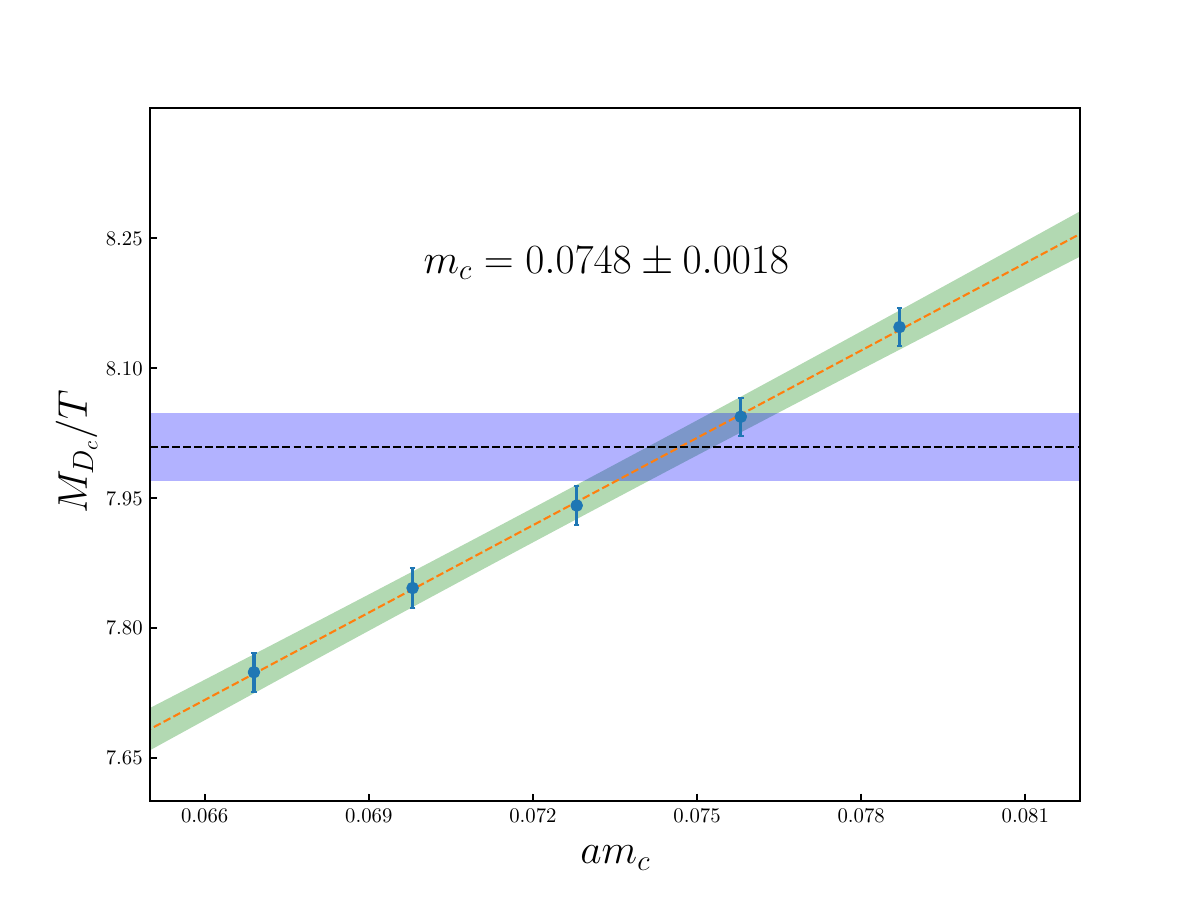}
  \caption{\label{fig:mq_tuning} Tuning of the charm quark mass $m_c$ at $N_\tau = 20$ and $T=511$ MeV.
  Left panel:  Effective mass of the correlation function for the $D_c$ meson (averaged over sources).
  The green band is the correlated two-exponential fit to the data.
  The data has clear correlations between datapoints, so the errors are computed via a bootstrap resampling of our configurations.
  Right panel: Matching of the fit to the $D_c$ correlator (green band) to the reference $D_c$ mass (blue band). The  overlap region provides the charm mass with uncertainties. }
\end{figure*}

\subsection{\label{sb:mq}Tuning of charm quark mass}
In tuning the charm quark mass,  we use a similar approach of trying several trial valence masses and comparing the resulting (thermal spatial-screening) meson masses to a reference meson mass.
We use the screening mass of the $D_s$ meson, because thermal effects obscure the influence of the light quark masses, and the $D_s$ is a flavor-nondiagonal charm-containing pseudoscalar.
The reference meson masses are measured on the reference configurations shown in bold in Table \ref{tab:reflats}.
In tuning the charm quark mass, we use the same gauge configurations generated in the confirmation run which uses an initial guess from the rational function approximation for the charm quark mass.
We then choose five quark masses in a 10\% mass range around this central value.
The strange quark masses are chosen correspondingly, using \Eq{massratios}.
We then compute the $D_s$ meson screening mass at each $m_c$ mass, scale them to $m_{D_s}/T$, and fit them to a linear function of $m_c$.
We then match this fit to the reference $m_{D_s}/T$ to determine the right physical $m_c$ and its error.
As a check, we perform the same procedure using the
unphysical $D_c$ meson, a $c\bar{c}$ state in which disconnected diagrams are ignored.
The operators for the $D_s$ and $D_c$ mesons in the continuum are :
\begin{align}
	D_s(x) &= \bar{c}^i_\alpha(x) \ (\gamma_5)_{\alpha \beta}  \ s^i_\beta(x)	\nonumber \\
	D_c(x) &= \bar{c}^i_\alpha(x) \ (\gamma_5)_{\alpha \beta}  \ c^i_\beta(x)
\end{align}
where the labels $c$ and $s$ correspond to the charm and strange spinors respectively.
Spin indices are in Greek and color indices are in Latin.
For the HISQ action, the spinor will be one component and the $\gamma_5$ will be replaced with the its staggered analogue. 
In computing the correlation functions, we employ a wall source in the screening direction and in order to guarantee sufficient statistics, we use twelve source slices spaced evenly in the screening direction.
We employ a two-exponential fit to account for contamination from higher mass states.
We illustrate the procedure in Figure \ref{fig:mq_tuning},
which shows the charm-quark measurement and the interpolation to the physical charm mass for the $N_\tau=20$, $\beta=7.82$ case.
We use bootstrap resampling to compute the errors, since this properly treats the rather large autocorrelations between the mass at nearby separations, which are visible in Figure \ref{fig:mq_tuning}.
The autocorrelations make it difficult to interpret a chisquare test, since this is usually expressed in terms of the number of \textsl{uncorrelated} degrees of freedom.

With the tuning of the charm quark mass and the coupling done, we perform a final matching run in order to confirm the final scale ratio.
The results are listed in Table \ref{tab:results}. 
We find that the $C_1,C_2$ ratios are indeed small and consistent with 0, and that the final uncertainties in the scale are within the desired tolerance of 1 to 2\%.

\begin{table*}[ht]
\centering
\caption{\label{tab:results} Final results of the scale setting including the tuning of bare coupling and charm quark  mass. }
 \renewcommand{\arraystretch}{1.5}
\begin{tabular}{|c|c|c|c|c|c|c|c|c|c|}
 \hline
 $\vphantom{\bigg |}$
Target $T$(MeV) &   $\beta_{\mathrm{tuned}}$ &  $am^{\mathrm{phys}}_c$ & $s$ & $10^4\, C_1$  & $10^4\, C_2$   & $\chi^2$/dof & $a$ (fm) &  $u_0$ & $T$(MeV) \\
\hline 
  \multirow{2}{*}{435(1)}  &  6.950 &  0.1994(13)  & 1.2496(71)  & -20(36) & 0.3(6) & 5/45 & 0.04531(26) & 0.89111(7) & 435(2)\\
    					   &  7.130 &  0.1636(13) & 1.4954(48)& -8(56) &  0.1(9) & 3.6/35& 0.03786(12) & 0.89508(4) & 434(2)\\ 
 \hline
  \multirow{3}{*}{341(2)} &  7.150  & 0.1432(30) & 0.8867(22) & -7(4) & 0.07(2) & 3.06/61 & 0.03626(9) & 0.89527(5) & 340(1)  \\
    					  &  7.390  & 0.1107(25) & 1.1173(40) & 4(50) & 0.0(4) & 5.81/75 & 0.02877(10) &  0.89981(3) & 342(1)\\
 	 			  		  &  7.715  & 0.0827(17) & 1.542(11)& 24(27)& 0.2(3) & 136/114&  0.02084(15)& 0.90528(1) &340(2)\\ 
 \hline
 \multirow{4}{*}{511(2)}   &  7.600   &  0.0884(45) &1.3394(23) & -14(4) & 0.2(3) & 8.3/49  & 0.02400(4) & 0.90310(5) & 513(1)\\
 					       &  7.820    & 0.0760(35) & 1.660(16) & -78(86) & 0.8(9) & 1.1/65 & 0.01936(19) & 0.90691(3) & 509(5)\\
 						   &  8.045   & 0.0697(46)  & 2.010(23) & 73(123) & 0(1) &  35/104 & 0.01599(18) & 0.91023(2) & 514(5)\\ 
 						   &  8.220    & 0.0559(17)  & 2.343(18) & -2(70) &  0.0(6) &  69/128 & 0.01372(11) & 0.91261(1)& 513(4) \\ 
 		\hline
 	\end{tabular}
 \end{table*}

\section{Conclusions and outlook}

In order to determine the topological susceptibility of QCD with precision to 1 GeV temperatures, we need accurate scale setting within a 2+1+1 flavor lattice gauge action which treats fermion chirality relatively well -- in our case, the HISQ framework -- with a precision of 1--2\% and down to lattice spacings of $a=0.014$ Fermi or $a^{-1} = 14\,\mathrm{GeV}$.
And we need a precise determination of the quark masses on these lattices.
We have done so by using step scaling, matching physical scales by matching gradient-flowed field strengths and matching quark masses by using $D_s$ meson masses.
Infrared issues are handled using thermal boundary conditions with temperatures between 341 and 511 MeV.
We have shown how to use thermal boundary conditions and to use a range of gradient-flow depths, simultaneously allowing precision and control of lattice-spacing effects.

\medskip

\begin{acknowledgments}
The authors acknowledge support by the Deutsche Forschungsgemeinschaft
(DFG, German Research Foundation) through the CRC-TR 211
``Strong-interaction matter under extreme conditions'' -- project
number 315477589 -- TRR 211. We also thank the GSI Helmholtzzentrum
and the TU Darmstadt and its Institut f\"ur Kernphysik for supporting
this research. Calculations were conducted on the Lichtenberg
high performance computer of the TU Darmstadt. This work was performed
using the framework of the publicly available MILC code \cite{MILC} and openQCD-1.6 package
\cite{openQCD}.
\end{acknowledgments}

\bibliography{refs}

\begin{thebibliography}{46}%
\makeatletter
\providecommand \@ifxundefined [1]{%
 \@ifx{#1\undefined}
}%
\providecommand \@ifnum [1]{%
 \ifnum #1\expandafter \@firstoftwo
 \else \expandafter \@secondoftwo
 \fi
}%
\providecommand \@ifx [1]{%
 \ifx #1\expandafter \@firstoftwo
 \else \expandafter \@secondoftwo
 \fi
}%
\providecommand \natexlab [1]{#1}%
\providecommand \enquote  [1]{``#1''}%
\providecommand \bibnamefont  [1]{#1}%
\providecommand \bibfnamefont [1]{#1}%
\providecommand \citenamefont [1]{#1}%
\providecommand \href@noop [0]{\@secondoftwo}%
\providecommand \href [0]{\begingroup \@sanitize@url \@href}%
\providecommand \@href[1]{\@@startlink{#1}\@@href}%
\providecommand \@@href[1]{\endgroup#1\@@endlink}%
\providecommand \@sanitize@url [0]{\catcode `\\12\catcode `\$12\catcode
  `\&12\catcode `\#12\catcode `\^12\catcode `\_12\catcode `\%12\relax}%
\providecommand \@@startlink[1]{}%
\providecommand \@@endlink[0]{}%
\providecommand \url  [0]{\begingroup\@sanitize@url \@url }%
\providecommand \@url [1]{\endgroup\@href {#1}{\urlprefix }}%
\providecommand \urlprefix  [0]{URL }%
\providecommand \Eprint [0]{\href }%
\providecommand \doibase [0]{http://dx.doi.org/}%
\providecommand \selectlanguage [0]{\@gobble}%
\providecommand \bibinfo  [0]{\@secondoftwo}%
\providecommand \bibfield  [0]{\@secondoftwo}%
\providecommand \translation [1]{[#1]}%
\providecommand \BibitemOpen [0]{}%
\providecommand \bibitemStop [0]{}%
\providecommand \bibitemNoStop [0]{.\EOS\space}%
\providecommand \EOS [0]{\spacefactor3000\relax}%
\providecommand \BibitemShut  [1]{\csname bibitem#1\endcsname}%
\let\auto@bib@innerbib\@empty
\bibitem [{\citenamefont {Weinberg}(1978)}]{Weinberg:1977ma}%
  \BibitemOpen
  \bibfield  {author} {\bibinfo {author} {\bibfnamefont {Steven}\ \bibnamefont
  {Weinberg}},\ }\bibfield  {title} {\enquote {\bibinfo {title} {{A New Light
  Boson?}}}\ }\href {\doibase 10.1103/PhysRevLett.40.223} {\bibfield  {journal}
  {\bibinfo  {journal} {Phys. Rev. Lett.}\ }\textbf {\bibinfo {volume} {40}},\
  \bibinfo {pages} {223--226} (\bibinfo {year} {1978})}\BibitemShut {NoStop}%
\bibitem [{\citenamefont {Wilczek}(1978)}]{Wilczek:1977pj}%
  \BibitemOpen
  \bibfield  {author} {\bibinfo {author} {\bibfnamefont {Frank}\ \bibnamefont
  {Wilczek}},\ }\bibfield  {title} {\enquote {\bibinfo {title} {{Problem of
  Strong $P$ and $T$ Invariance in the Presence of Instantons}},}\ }\href
  {\doibase 10.1103/PhysRevLett.40.279} {\bibfield  {journal} {\bibinfo
  {journal} {Phys. Rev. Lett.}\ }\textbf {\bibinfo {volume} {40}},\ \bibinfo
  {pages} {279--282} (\bibinfo {year} {1978})}\BibitemShut {NoStop}%
\bibitem [{\citenamefont {Preskill}\ \emph {et~al.}(1983)\citenamefont
  {Preskill}, \citenamefont {Wise},\ and\ \citenamefont
  {Wilczek}}]{Preskill:1982cy}%
  \BibitemOpen
  \bibfield  {author} {\bibinfo {author} {\bibfnamefont {John}\ \bibnamefont
  {Preskill}}, \bibinfo {author} {\bibfnamefont {Mark~B.}\ \bibnamefont
  {Wise}}, \ and\ \bibinfo {author} {\bibfnamefont {Frank}\ \bibnamefont
  {Wilczek}},\ }\bibfield  {title} {\enquote {\bibinfo {title} {{Cosmology of
  the Invisible Axion}},}\ }\href {\doibase 10.1016/0370-2693(83)90637-8}
  {\bibfield  {journal} {\bibinfo  {journal} {Phys. Lett. B}\ }\textbf
  {\bibinfo {volume} {120}},\ \bibinfo {pages} {127--132} (\bibinfo {year}
  {1983})}\BibitemShut {NoStop}%
\bibitem [{\citenamefont {Abbott}\ and\ \citenamefont
  {Sikivie}(1983)}]{Abbott:1982af}%
  \BibitemOpen
  \bibfield  {author} {\bibinfo {author} {\bibfnamefont {L.~F.}\ \bibnamefont
  {Abbott}}\ and\ \bibinfo {author} {\bibfnamefont {P.}~\bibnamefont
  {Sikivie}},\ }\bibfield  {title} {\enquote {\bibinfo {title} {{A Cosmological
  Bound on the Invisible Axion}},}\ }\href {\doibase
  10.1016/0370-2693(83)90638-X} {\bibfield  {journal} {\bibinfo  {journal}
  {Phys. Lett. B}\ }\textbf {\bibinfo {volume} {120}},\ \bibinfo {pages}
  {133--136} (\bibinfo {year} {1983})}\BibitemShut {NoStop}%
\bibitem [{\citenamefont {Dine}\ and\ \citenamefont
  {Fischler}(1983)}]{Dine:1982ah}%
  \BibitemOpen
  \bibfield  {author} {\bibinfo {author} {\bibfnamefont {Michael}\ \bibnamefont
  {Dine}}\ and\ \bibinfo {author} {\bibfnamefont {Willy}\ \bibnamefont
  {Fischler}},\ }\bibfield  {title} {\enquote {\bibinfo {title} {{The Not So
  Harmless Axion}},}\ }\href {\doibase 10.1016/0370-2693(83)90639-1} {\bibfield
   {journal} {\bibinfo  {journal} {Phys. Lett. B}\ }\textbf {\bibinfo {volume}
  {120}},\ \bibinfo {pages} {137--141} (\bibinfo {year} {1983})}\BibitemShut
  {NoStop}%
\bibitem [{\citenamefont {Visinelli}\ and\ \citenamefont
  {Gondolo}(2009)}]{Visinelli:2009zm}%
  \BibitemOpen
  \bibfield  {author} {\bibinfo {author} {\bibfnamefont {Luca}\ \bibnamefont
  {Visinelli}}\ and\ \bibinfo {author} {\bibfnamefont {Paolo}\ \bibnamefont
  {Gondolo}},\ }\bibfield  {title} {\enquote {\bibinfo {title} {{Dark Matter
  Axions Revisited}},}\ }\href {\doibase 10.1103/PhysRevD.80.035024} {\bibfield
   {journal} {\bibinfo  {journal} {Phys. Rev. D}\ }\textbf {\bibinfo {volume}
  {80}},\ \bibinfo {pages} {035024} (\bibinfo {year} {2009})},\ \Eprint
  {http://arxiv.org/abs/0903.4377} {arXiv:0903.4377 [astro-ph.CO]} \BibitemShut
  {NoStop}%
\bibitem [{\citenamefont {Klaer}\ and\ \citenamefont
  {Moore}(2017)}]{Klaer:2017qhr}%
  \BibitemOpen
  \bibfield  {author} {\bibinfo {author} {\bibfnamefont {Vincent~B.}\
  \bibnamefont {Klaer}}\ and\ \bibinfo {author} {\bibfnamefont {Guy~D.}\
  \bibnamefont {Moore}},\ }\bibfield  {title} {\enquote {\bibinfo {title} {{How
  to simulate global cosmic strings with large string tension}},}\ }\href
  {\doibase 10.1088/1475-7516/2017/10/043} {\bibfield  {journal} {\bibinfo
  {journal} {JCAP}\ }\textbf {\bibinfo {volume} {10}},\ \bibinfo {pages} {043}
  (\bibinfo {year} {2017})},\ \Eprint {http://arxiv.org/abs/1707.05566}
  {arXiv:1707.05566 [hep-ph]} \BibitemShut {NoStop}%
\bibitem [{\citenamefont {Vaquero}\ \emph {et~al.}(2019)\citenamefont
  {Vaquero}, \citenamefont {Redondo},\ and\ \citenamefont
  {Stadler}}]{Vaquero:2018tib}%
  \BibitemOpen
  \bibfield  {author} {\bibinfo {author} {\bibfnamefont {Alejandro}\
  \bibnamefont {Vaquero}}, \bibinfo {author} {\bibfnamefont {Javier}\
  \bibnamefont {Redondo}}, \ and\ \bibinfo {author} {\bibfnamefont {Julia}\
  \bibnamefont {Stadler}},\ }\bibfield  {title} {\enquote {\bibinfo {title}
  {{Early seeds of axion miniclusters}},}\ }\href {\doibase
  10.1088/1475-7516/2019/04/012} {\bibfield  {journal} {\bibinfo  {journal}
  {JCAP}\ }\textbf {\bibinfo {volume} {04}},\ \bibinfo {pages} {012} (\bibinfo
  {year} {2019})},\ \Eprint {http://arxiv.org/abs/1809.09241} {arXiv:1809.09241
  [astro-ph.CO]} \BibitemShut {NoStop}%
\bibitem [{\citenamefont {Buschmann}\ \emph {et~al.}(2020)\citenamefont
  {Buschmann}, \citenamefont {Foster},\ and\ \citenamefont
  {Safdi}}]{Buschmann:2019icd}%
  \BibitemOpen
  \bibfield  {author} {\bibinfo {author} {\bibfnamefont {Malte}\ \bibnamefont
  {Buschmann}}, \bibinfo {author} {\bibfnamefont {Joshua~W.}\ \bibnamefont
  {Foster}}, \ and\ \bibinfo {author} {\bibfnamefont {Benjamin~R.}\
  \bibnamefont {Safdi}},\ }\bibfield  {title} {\enquote {\bibinfo {title}
  {{Early-Universe Simulations of the Cosmological Axion}},}\ }\href {\doibase
  10.1103/PhysRevLett.124.161103} {\bibfield  {journal} {\bibinfo  {journal}
  {Phys. Rev. Lett.}\ }\textbf {\bibinfo {volume} {124}},\ \bibinfo {pages}
  {161103} (\bibinfo {year} {2020})},\ \Eprint
  {http://arxiv.org/abs/1906.00967} {arXiv:1906.00967 [astro-ph.CO]}
  \BibitemShut {NoStop}%
\bibitem [{\citenamefont {Moore}(2018)}]{Moore:2017ond}%
  \BibitemOpen
  \bibfield  {author} {\bibinfo {author} {\bibfnamefont {Guy~D.}\ \bibnamefont
  {Moore}},\ }\bibfield  {title} {\enquote {\bibinfo {title} {{Axion dark
  matter and the Lattice}},}\ }\href {\doibase 10.1051/epjconf/201817501009}
  {\bibfield  {journal} {\bibinfo  {journal} {EPJ Web Conf.}\ }\textbf
  {\bibinfo {volume} {175}},\ \bibinfo {pages} {01009} (\bibinfo {year}
  {2018})},\ \Eprint {http://arxiv.org/abs/1709.09466} {arXiv:1709.09466
  [hep-ph]} \BibitemShut {NoStop}%
\bibitem [{\citenamefont {Petreczky}\ \emph {et~al.}(2016)\citenamefont
  {Petreczky}, \citenamefont {Schadler},\ and\ \citenamefont
  {Sharma}}]{Petreczky:2016vrs}%
  \BibitemOpen
  \bibfield  {author} {\bibinfo {author} {\bibfnamefont {Peter}\ \bibnamefont
  {Petreczky}}, \bibinfo {author} {\bibfnamefont {Hans-Peter}\ \bibnamefont
  {Schadler}}, \ and\ \bibinfo {author} {\bibfnamefont {Sayantan}\ \bibnamefont
  {Sharma}},\ }\bibfield  {title} {\enquote {\bibinfo {title} {{The topological
  susceptibility in finite temperature QCD and axion cosmology}},}\ }\href
  {\doibase 10.1016/j.physletb.2016.09.063} {\bibfield  {journal} {\bibinfo
  {journal} {Phys. Lett. B}\ }\textbf {\bibinfo {volume} {762}},\ \bibinfo
  {pages} {498--505} (\bibinfo {year} {2016})},\ \Eprint
  {http://arxiv.org/abs/1606.03145} {arXiv:1606.03145 [hep-lat]} \BibitemShut
  {NoStop}%
\bibitem [{\citenamefont {Bonati}\ \emph {et~al.}(2018)\citenamefont {Bonati},
  \citenamefont {D'Elia}, \citenamefont {Martinelli}, \citenamefont {Negro},
  \citenamefont {Sanfilippo},\ and\ \citenamefont {Todaro}}]{Bonati:2018blm}%
  \BibitemOpen
  \bibfield  {author} {\bibinfo {author} {\bibfnamefont {Claudio}\ \bibnamefont
  {Bonati}}, \bibinfo {author} {\bibfnamefont {Massimo}\ \bibnamefont
  {D'Elia}}, \bibinfo {author} {\bibfnamefont {Guido}\ \bibnamefont
  {Martinelli}}, \bibinfo {author} {\bibfnamefont {Francesco}\ \bibnamefont
  {Negro}}, \bibinfo {author} {\bibfnamefont {Francesco}\ \bibnamefont
  {Sanfilippo}}, \ and\ \bibinfo {author} {\bibfnamefont {Antonino}\
  \bibnamefont {Todaro}},\ }\bibfield  {title} {\enquote {\bibinfo {title}
  {{Topology in full QCD at high temperature: a multicanonical approach}},}\
  }\href {\doibase 10.1007/JHEP11(2018)170} {\bibfield  {journal} {\bibinfo
  {journal} {JHEP}\ }\textbf {\bibinfo {volume} {11}},\ \bibinfo {pages} {170}
  (\bibinfo {year} {2018})},\ \Eprint {http://arxiv.org/abs/1807.07954}
  {arXiv:1807.07954 [hep-lat]} \BibitemShut {NoStop}%
\bibitem [{\citenamefont {Athenodorou}\ \emph {et~al.}(2022)\citenamefont
  {Athenodorou}, \citenamefont {Bonanno}, \citenamefont {Bonati}, \citenamefont
  {Clemente}, \citenamefont {D'Angelo}, \citenamefont {D'Elia}, \citenamefont
  {Maio}, \citenamefont {Martinelli}, \citenamefont {Sanfilippo},\ and\
  \citenamefont {Todaro}}]{Athenodorou:2022aay}%
  \BibitemOpen
  \bibfield  {author} {\bibinfo {author} {\bibfnamefont {Andreas}\ \bibnamefont
  {Athenodorou}}, \bibinfo {author} {\bibfnamefont {Claudio}\ \bibnamefont
  {Bonanno}}, \bibinfo {author} {\bibfnamefont {Claudio}\ \bibnamefont
  {Bonati}}, \bibinfo {author} {\bibfnamefont {Giuseppe}\ \bibnamefont
  {Clemente}}, \bibinfo {author} {\bibfnamefont {Francesco}\ \bibnamefont
  {D'Angelo}}, \bibinfo {author} {\bibfnamefont {Massimo}\ \bibnamefont
  {D'Elia}}, \bibinfo {author} {\bibfnamefont {Lorenzo}\ \bibnamefont {Maio}},
  \bibinfo {author} {\bibfnamefont {Guido}\ \bibnamefont {Martinelli}},
  \bibinfo {author} {\bibfnamefont {Francesco}\ \bibnamefont {Sanfilippo}}, \
  and\ \bibinfo {author} {\bibfnamefont {Antonino}\ \bibnamefont {Todaro}},\
  }\bibfield  {title} {\enquote {\bibinfo {title} {{Topological susceptibility
  of N$_{f}$ = 2 + 1 QCD from staggered fermions spectral projectors at high
  temperatures}},}\ }\href {\doibase 10.1007/JHEP10(2022)197} {\bibfield
  {journal} {\bibinfo  {journal} {JHEP}\ }\textbf {\bibinfo {volume} {10}},\
  \bibinfo {pages} {197} (\bibinfo {year} {2022})},\ \Eprint
  {http://arxiv.org/abs/2208.08921} {arXiv:2208.08921 [hep-lat]} \BibitemShut
  {NoStop}%
\bibitem [{\citenamefont {Borsanyi}\ \emph {et~al.}(2016)\citenamefont
  {Borsanyi} \emph {et~al.}}]{Borsanyi:2016ksw}%
  \BibitemOpen
  \bibfield  {author} {\bibinfo {author} {\bibfnamefont {Sz.}\ \bibnamefont
  {Borsanyi}} \emph {et~al.},\ }\bibfield  {title} {\enquote {\bibinfo {title}
  {{Calculation of the axion mass based on high-temperature lattice quantum
  chromodynamics}},}\ }\href {\doibase 10.1038/nature20115} {\bibfield
  {journal} {\bibinfo  {journal} {Nature}\ }\textbf {\bibinfo {volume} {539}},\
  \bibinfo {pages} {69--71} (\bibinfo {year} {2016})},\ \Eprint
  {http://arxiv.org/abs/1606.07494} {arXiv:1606.07494 [hep-lat]} \BibitemShut
  {NoStop}%
\bibitem [{\citenamefont {Jahn}\ \emph {et~al.}(2018)\citenamefont {Jahn},
  \citenamefont {Moore},\ and\ \citenamefont {Robaina}}]{Jahn:2018dke}%
  \BibitemOpen
  \bibfield  {author} {\bibinfo {author} {\bibfnamefont {Peter~Thomas}\
  \bibnamefont {Jahn}}, \bibinfo {author} {\bibfnamefont {Guy~D.}\ \bibnamefont
  {Moore}}, \ and\ \bibinfo {author} {\bibfnamefont {Daniel}\ \bibnamefont
  {Robaina}},\ }\bibfield  {title} {\enquote {\bibinfo {title}
  {{$\chi_{\textrm{top}}(T \gg T_{\textrm{c}})$ in pure-glue QCD through
  reweighting}},}\ }\href {\doibase 10.1103/PhysRevD.98.054512} {\bibfield
  {journal} {\bibinfo  {journal} {Phys. Rev. D}\ }\textbf {\bibinfo {volume}
  {98}},\ \bibinfo {pages} {054512} (\bibinfo {year} {2018})},\ \Eprint
  {http://arxiv.org/abs/1806.01162} {arXiv:1806.01162 [hep-lat]} \BibitemShut
  {NoStop}%
\bibitem [{\citenamefont {Jahn}\ \emph {et~al.}(2020)\citenamefont {Jahn},
  \citenamefont {Moore},\ and\ \citenamefont {Robaina}}]{Jahn:2020oqf}%
  \BibitemOpen
  \bibfield  {author} {\bibinfo {author} {\bibfnamefont {P.~Thomas}\
  \bibnamefont {Jahn}}, \bibinfo {author} {\bibfnamefont {Guy~D.}\ \bibnamefont
  {Moore}}, \ and\ \bibinfo {author} {\bibfnamefont {Daniel}\ \bibnamefont
  {Robaina}},\ }\bibfield  {title} {\enquote {\bibinfo {title} {{Improved
  Reweighting for QCD Topology at High Temperature}},}\ }\href@noop {} {\
  (\bibinfo {year} {2020})},\ \Eprint {http://arxiv.org/abs/2002.01153}
  {arXiv:2002.01153 [hep-lat]} \BibitemShut {NoStop}%
\bibitem [{\citenamefont {Gross}\ \emph {et~al.}(1981)\citenamefont {Gross},
  \citenamefont {Pisarski},\ and\ \citenamefont {Yaffe}}]{Gross:1980br}%
  \BibitemOpen
  \bibfield  {author} {\bibinfo {author} {\bibfnamefont {David~J.}\
  \bibnamefont {Gross}}, \bibinfo {author} {\bibfnamefont {Robert~D.}\
  \bibnamefont {Pisarski}}, \ and\ \bibinfo {author} {\bibfnamefont
  {Laurence~G.}\ \bibnamefont {Yaffe}},\ }\bibfield  {title} {\enquote
  {\bibinfo {title} {{QCD and Instantons at Finite Temperature}},}\ }\href
  {\doibase 10.1103/RevModPhys.53.43} {\bibfield  {journal} {\bibinfo
  {journal} {Rev. Mod. Phys.}\ }\textbf {\bibinfo {volume} {53}},\ \bibinfo
  {pages} {43} (\bibinfo {year} {1981})}\BibitemShut {NoStop}%
\bibitem [{\citenamefont {Kanazawa}\ and\ \citenamefont
  {Yamamoto}(2015)}]{Kanazawa:2014cua}%
  \BibitemOpen
  \bibfield  {author} {\bibinfo {author} {\bibfnamefont {Takuya}\ \bibnamefont
  {Kanazawa}}\ and\ \bibinfo {author} {\bibfnamefont {Naoki}\ \bibnamefont
  {Yamamoto}},\ }\bibfield  {title} {\enquote {\bibinfo {title}
  {{Quasi-instantons in QCD with chiral symmetry restoration}},}\ }\href
  {\doibase 10.1103/PhysRevD.91.105015} {\bibfield  {journal} {\bibinfo
  {journal} {Phys. Rev. D}\ }\textbf {\bibinfo {volume} {91}},\ \bibinfo
  {pages} {105015} (\bibinfo {year} {2015})},\ \Eprint
  {http://arxiv.org/abs/1410.3614} {arXiv:1410.3614 [hep-th]} \BibitemShut
  {NoStop}%
\bibitem [{\citenamefont {Follana}\ \emph {et~al.}(2007)\citenamefont
  {Follana}, \citenamefont {Mason}, \citenamefont {Davies}, \citenamefont
  {Hornbostel}, \citenamefont {Lepage}, \citenamefont {Shigemitsu},
  \citenamefont {Trottier},\ and\ \citenamefont {Wong}}]{Follana:2006rc}%
  \BibitemOpen
  \bibfield  {author} {\bibinfo {author} {\bibfnamefont {E.}~\bibnamefont
  {Follana}}, \bibinfo {author} {\bibfnamefont {Q.}~\bibnamefont {Mason}},
  \bibinfo {author} {\bibfnamefont {C.}~\bibnamefont {Davies}}, \bibinfo
  {author} {\bibfnamefont {K.}~\bibnamefont {Hornbostel}}, \bibinfo {author}
  {\bibfnamefont {G.~P.}\ \bibnamefont {Lepage}}, \bibinfo {author}
  {\bibfnamefont {J.}~\bibnamefont {Shigemitsu}}, \bibinfo {author}
  {\bibfnamefont {H.}~\bibnamefont {Trottier}}, \ and\ \bibinfo {author}
  {\bibfnamefont {K.}~\bibnamefont {Wong}} (\bibinfo {collaboration} {HPQCD,
  UKQCD}),\ }\bibfield  {title} {\enquote {\bibinfo {title} {{Highly improved
  staggered quarks on the lattice, with applications to charm physics}},}\
  }\href {\doibase 10.1103/PhysRevD.75.054502} {\bibfield  {journal} {\bibinfo
  {journal} {Phys. Rev. D}\ }\textbf {\bibinfo {volume} {75}},\ \bibinfo
  {pages} {054502} (\bibinfo {year} {2007})},\ \Eprint
  {http://arxiv.org/abs/hep-lat/0610092} {arXiv:hep-lat/0610092} \BibitemShut
  {NoStop}%
\bibitem [{\citenamefont {Sommer}(1994)}]{Sommer:1993ce}%
  \BibitemOpen
  \bibfield  {author} {\bibinfo {author} {\bibfnamefont {R.}~\bibnamefont
  {Sommer}},\ }\bibfield  {title} {\enquote {\bibinfo {title} {{A New way to
  set the energy scale in lattice gauge theories and its applications to the
  static force and alpha-s in SU(2) Yang-Mills theory}},}\ }\href {\doibase
  10.1016/0550-3213(94)90473-1} {\bibfield  {journal} {\bibinfo  {journal}
  {Nucl. Phys. B}\ }\textbf {\bibinfo {volume} {411}},\ \bibinfo {pages}
  {839--854} (\bibinfo {year} {1994})},\ \Eprint
  {http://arxiv.org/abs/hep-lat/9310022} {arXiv:hep-lat/9310022} \BibitemShut
  {NoStop}%
\bibitem [{\citenamefont {Bazavov}\ \emph {et~al.}(2013)\citenamefont {Bazavov}
  \emph {et~al.}}]{MILC:2012znn}%
  \BibitemOpen
  \bibfield  {author} {\bibinfo {author} {\bibfnamefont {A.}~\bibnamefont
  {Bazavov}} \emph {et~al.} (\bibinfo {collaboration} {MILC}),\ }\bibfield
  {title} {\enquote {\bibinfo {title} {{Lattice QCD Ensembles with Four Flavors
  of Highly Improved Staggered Quarks}},}\ }\href {\doibase
  10.1103/PhysRevD.87.054505} {\bibfield  {journal} {\bibinfo  {journal} {Phys.
  Rev. D}\ }\textbf {\bibinfo {volume} {87}},\ \bibinfo {pages} {054505}
  (\bibinfo {year} {2013})},\ \Eprint {http://arxiv.org/abs/1212.4768}
  {arXiv:1212.4768 [hep-lat]} \BibitemShut {NoStop}%
\bibitem [{\citenamefont {Bazavov}\ \emph {et~al.}(2018)\citenamefont {Bazavov}
  \emph {et~al.}}]{Bazavov:2017lyh}%
  \BibitemOpen
  \bibfield  {author} {\bibinfo {author} {\bibfnamefont {A.}~\bibnamefont
  {Bazavov}} \emph {et~al.},\ }\bibfield  {title} {\enquote {\bibinfo {title}
  {{$B$- and $D$-meson leptonic decay constants from four-flavor lattice
  QCD}},}\ }\href {\doibase 10.1103/PhysRevD.98.074512} {\bibfield  {journal}
  {\bibinfo  {journal} {Phys. Rev. D}\ }\textbf {\bibinfo {volume} {98}},\
  \bibinfo {pages} {074512} (\bibinfo {year} {2018})},\ \Eprint
  {http://arxiv.org/abs/1712.09262} {arXiv:1712.09262 [hep-lat]} \BibitemShut
  {NoStop}%
\bibitem [{\citenamefont {Luscher}\ \emph {et~al.}(1991)\citenamefont
  {Luscher}, \citenamefont {Weisz},\ and\ \citenamefont
  {Wolff}}]{Luscher:1991wu}%
  \BibitemOpen
  \bibfield  {author} {\bibinfo {author} {\bibfnamefont {Martin}\ \bibnamefont
  {Luscher}}, \bibinfo {author} {\bibfnamefont {Peter}\ \bibnamefont {Weisz}},
  \ and\ \bibinfo {author} {\bibfnamefont {Ulli}\ \bibnamefont {Wolff}},\
  }\bibfield  {title} {\enquote {\bibinfo {title} {{A Numerical method to
  compute the running coupling in asymptotically free theories}},}\ }\href
  {\doibase 10.1016/0550-3213(91)90298-C} {\bibfield  {journal} {\bibinfo
  {journal} {Nucl. Phys. B}\ }\textbf {\bibinfo {volume} {359}},\ \bibinfo
  {pages} {221--243} (\bibinfo {year} {1991})}\BibitemShut {NoStop}%
\bibitem [{\citenamefont {Nada}\ and\ \citenamefont
  {Ramos}(2021)}]{Nada:2020jay}%
  \BibitemOpen
  \bibfield  {author} {\bibinfo {author} {\bibfnamefont {Alessandro}\
  \bibnamefont {Nada}}\ and\ \bibinfo {author} {\bibfnamefont {Alberto}\
  \bibnamefont {Ramos}},\ }\bibfield  {title} {\enquote {\bibinfo {title} {{An
  analysis of systematic effects in finite size scaling studies using the
  gradient flow}},}\ }\href {\doibase 10.1140/epjc/s10052-020-08759-1}
  {\bibfield  {journal} {\bibinfo  {journal} {Eur. Phys. J. C}\ }\textbf
  {\bibinfo {volume} {81}},\ \bibinfo {pages} {1} (\bibinfo {year} {2021})},\
  \Eprint {http://arxiv.org/abs/2007.12862} {arXiv:2007.12862 [hep-lat]}
  \BibitemShut {NoStop}%
\bibitem [{\citenamefont {Narayanan}\ and\ \citenamefont
  {Neuberger}(2006)}]{Narayanan:2006rf}%
  \BibitemOpen
  \bibfield  {author} {\bibinfo {author} {\bibfnamefont {R.}~\bibnamefont
  {Narayanan}}\ and\ \bibinfo {author} {\bibfnamefont {H.}~\bibnamefont
  {Neuberger}},\ }\bibfield  {title} {\enquote {\bibinfo {title} {{Infinite N
  phase transitions in continuum Wilson loop operators}},}\ }\href {\doibase
  10.1088/1126-6708/2006/03/064} {\bibfield  {journal} {\bibinfo  {journal}
  {JHEP}\ }\textbf {\bibinfo {volume} {03}},\ \bibinfo {pages} {064} (\bibinfo
  {year} {2006})},\ \Eprint {http://arxiv.org/abs/hep-th/0601210}
  {arXiv:hep-th/0601210} \BibitemShut {NoStop}%
\bibitem [{\citenamefont {L\"uscher}(2010)}]{Luscher:2010iy}%
  \BibitemOpen
  \bibfield  {author} {\bibinfo {author} {\bibfnamefont {Martin}\ \bibnamefont
  {L\"uscher}},\ }\bibfield  {title} {\enquote {\bibinfo {title} {{Properties
  and uses of the Wilson flow in lattice QCD}},}\ }\href {\doibase
  10.1007/JHEP08(2010)071} {\bibfield  {journal} {\bibinfo  {journal} {JHEP}\
  }\textbf {\bibinfo {volume} {08}},\ \bibinfo {pages} {071} (\bibinfo {year}
  {2010})},\ \bibinfo {note} {[Erratum: JHEP 03, 092 (2014)]},\ \Eprint
  {http://arxiv.org/abs/1006.4518} {arXiv:1006.4518 [hep-lat]} \BibitemShut
  {NoStop}%
\bibitem [{\citenamefont {Gross}\ and\ \citenamefont
  {Wilczek}(1973)}]{Gross:1973id}%
  \BibitemOpen
  \bibfield  {author} {\bibinfo {author} {\bibfnamefont {David~J.}\
  \bibnamefont {Gross}}\ and\ \bibinfo {author} {\bibfnamefont {Frank}\
  \bibnamefont {Wilczek}},\ }\bibfield  {title} {\enquote {\bibinfo {title}
  {{Ultraviolet Behavior of Nonabelian Gauge Theories}},}\ }\href {\doibase
  10.1103/PhysRevLett.30.1343} {\bibfield  {journal} {\bibinfo  {journal}
  {Phys. Rev. Lett.}\ }\textbf {\bibinfo {volume} {30}},\ \bibinfo {pages}
  {1343--1346} (\bibinfo {year} {1973})}\BibitemShut {NoStop}%
\bibitem [{\citenamefont {Politzer}(1973)}]{Politzer:1973fx}%
  \BibitemOpen
  \bibfield  {author} {\bibinfo {author} {\bibfnamefont {H.~David}\
  \bibnamefont {Politzer}},\ }\bibfield  {title} {\enquote {\bibinfo {title}
  {{Reliable Perturbative Results for Strong Interactions?}}}\ }\href {\doibase
  10.1103/PhysRevLett.30.1346} {\bibfield  {journal} {\bibinfo  {journal}
  {Phys. Rev. Lett.}\ }\textbf {\bibinfo {volume} {30}},\ \bibinfo {pages}
  {1346--1349} (\bibinfo {year} {1973})}\BibitemShut {NoStop}%
\bibitem [{\citenamefont {Luscher}\ and\ \citenamefont
  {Weisz}(2011)}]{Luscher:2011bx}%
  \BibitemOpen
  \bibfield  {author} {\bibinfo {author} {\bibfnamefont {Martin}\ \bibnamefont
  {Luscher}}\ and\ \bibinfo {author} {\bibfnamefont {Peter}\ \bibnamefont
  {Weisz}},\ }\bibfield  {title} {\enquote {\bibinfo {title} {{Perturbative
  Analysis of the Gradient Flow in Non-Abelian Gauge Theories}},}\ }\href
  {\doibase 10.1007/JHEP02(2011)051} {\bibfield  {journal} {\bibinfo  {journal}
  {JHEP}\ }\textbf {\bibinfo {volume} {02}},\ \bibinfo {pages} {051} (\bibinfo
  {year} {2011})},\ \Eprint {http://arxiv.org/abs/1101.0963} {arXiv:1101.0963
  [hep-th]} \BibitemShut {NoStop}%
\bibitem [{\citenamefont {Harlander}\ and\ \citenamefont
  {Neumann}(2016)}]{Harlander:2016vzb}%
  \BibitemOpen
  \bibfield  {author} {\bibinfo {author} {\bibfnamefont {Robert~V.}\
  \bibnamefont {Harlander}}\ and\ \bibinfo {author} {\bibfnamefont {Tobias}\
  \bibnamefont {Neumann}},\ }\bibfield  {title} {\enquote {\bibinfo {title}
  {{The perturbative QCD gradient flow to three loops}},}\ }\href {\doibase
  10.1007/JHEP06(2016)161} {\bibfield  {journal} {\bibinfo  {journal} {JHEP}\
  }\textbf {\bibinfo {volume} {06}},\ \bibinfo {pages} {161} (\bibinfo {year}
  {2016})},\ \Eprint {http://arxiv.org/abs/1606.03756} {arXiv:1606.03756
  [hep-ph]} \BibitemShut {NoStop}%
\bibitem [{\citenamefont {Borsanyi}\ \emph {et~al.}(2012)\citenamefont
  {Borsanyi} \emph {et~al.}}]{BMW:2012hcm}%
  \BibitemOpen
  \bibfield  {author} {\bibinfo {author} {\bibfnamefont {Szabolcs}\
  \bibnamefont {Borsanyi}} \emph {et~al.} (\bibinfo {collaboration} {BMW}),\
  }\bibfield  {title} {\enquote {\bibinfo {title} {{High-precision scale
  setting in lattice QCD}},}\ }\href {\doibase 10.1007/JHEP09(2012)010}
  {\bibfield  {journal} {\bibinfo  {journal} {JHEP}\ }\textbf {\bibinfo
  {volume} {09}},\ \bibinfo {pages} {010} (\bibinfo {year} {2012})},\ \Eprint
  {http://arxiv.org/abs/1203.4469} {arXiv:1203.4469 [hep-lat]} \BibitemShut
  {NoStop}%
\bibitem [{\citenamefont {Fodor}\ \emph {et~al.}(2014)\citenamefont {Fodor},
  \citenamefont {Holland}, \citenamefont {Kuti}, \citenamefont {Mondal},
  \citenamefont {Nogradi},\ and\ \citenamefont {Wong}}]{Fodor:2014cpa}%
  \BibitemOpen
  \bibfield  {author} {\bibinfo {author} {\bibfnamefont {Zoltan}\ \bibnamefont
  {Fodor}}, \bibinfo {author} {\bibfnamefont {Kieran}\ \bibnamefont {Holland}},
  \bibinfo {author} {\bibfnamefont {Julius}\ \bibnamefont {Kuti}}, \bibinfo
  {author} {\bibfnamefont {Santanu}\ \bibnamefont {Mondal}}, \bibinfo {author}
  {\bibfnamefont {Daniel}\ \bibnamefont {Nogradi}}, \ and\ \bibinfo {author}
  {\bibfnamefont {Chik~Him}\ \bibnamefont {Wong}},\ }\bibfield  {title}
  {\enquote {\bibinfo {title} {{The Lattice Gradient Flow at Tree-Level and Its
  Improvement}},}\ }\href {\doibase 10.1007/JHEP09(2014)018} {\bibfield
  {journal} {\bibinfo  {journal} {JHEP}\ }\textbf {\bibinfo {volume} {09}},\
  \bibinfo {pages} {018} (\bibinfo {year} {2014})},\ \Eprint
  {http://arxiv.org/abs/1406.0827} {arXiv:1406.0827 [hep-lat]} \BibitemShut
  {NoStop}%
\bibitem [{\citenamefont {Luscher}\ and\ \citenamefont
  {Weisz}(1985)}]{Luscher:1984xn}%
  \BibitemOpen
  \bibfield  {author} {\bibinfo {author} {\bibfnamefont {M.}~\bibnamefont
  {Luscher}}\ and\ \bibinfo {author} {\bibfnamefont {P.}~\bibnamefont
  {Weisz}},\ }\bibfield  {title} {\enquote {\bibinfo {title} {{On-Shell
  Improved Lattice Gauge Theories}},}\ }\href {\doibase 10.1007/BF01206178}
  {\bibfield  {journal} {\bibinfo  {journal} {Commun. Math. Phys.}\ }\textbf
  {\bibinfo {volume} {97}},\ \bibinfo {pages} {59} (\bibinfo {year} {1985})},\
  \bibinfo {note} {[Erratum: Commun.Math.Phys. 98, 433 (1985)]}\BibitemShut
  {NoStop}%
\bibitem [{\citenamefont {Ramos}\ and\ \citenamefont
  {Sint}(2016)}]{Ramos:2015baa}%
  \BibitemOpen
  \bibfield  {author} {\bibinfo {author} {\bibfnamefont {A.}~\bibnamefont
  {Ramos}}\ and\ \bibinfo {author} {\bibfnamefont {S.}~\bibnamefont {Sint}},\
  }\bibfield  {title} {\enquote {\bibinfo {title} {{Symanzik Improvement of the
  Gradient Flow in Lattice Gauge Theories}},}\ }\href {\doibase
  10.1140/epjc/s10052-015-3831-9} {\bibfield  {journal} {\bibinfo  {journal}
  {Eur. Phys. J. C}\ }\textbf {\bibinfo {volume} {76}},\ \bibinfo {pages} {15}
  (\bibinfo {year} {2016})},\ \Eprint {http://arxiv.org/abs/1508.05552}
  {arXiv:1508.05552 [hep-lat]} \BibitemShut {NoStop}%
\bibitem [{\citenamefont {Stendebach}(2023)}]{tuprints23185}%
  \BibitemOpen
  \bibfield  {author} {\bibinfo {author} {\bibfnamefont {Simon}\ \bibnamefont
  {Stendebach}},\ }\emph {\bibinfo {title} {Perturbative analysis of operators
  under improved gradient flow in lattice QCD}},\ \href {\doibase
  https://doi.org/10.26083/tuprints-00023185} {Master's thesis},\ \bibinfo
  {school} {Technische Universit{\"a}t Darmstadt}, \bibinfo {address}
  {Darmstadt} (\bibinfo {year} {2023})\BibitemShut {NoStop}%
\bibitem [{\citenamefont {Eller}\ and\ \citenamefont
  {Moore}(2018)}]{Eller:2018yje}%
  \BibitemOpen
  \bibfield  {author} {\bibinfo {author} {\bibfnamefont {Alexander~M.}\
  \bibnamefont {Eller}}\ and\ \bibinfo {author} {\bibfnamefont {Guy~D.}\
  \bibnamefont {Moore}},\ }\bibfield  {title} {\enquote {\bibinfo {title}
  {{Gradient-flowed thermal correlators: how much flow is too much?}}}\ }\href
  {\doibase 10.1103/PhysRevD.97.114507} {\bibfield  {journal} {\bibinfo
  {journal} {Phys. Rev. D}\ }\textbf {\bibinfo {volume} {97}},\ \bibinfo
  {pages} {114507} (\bibinfo {year} {2018})},\ \Eprint
  {http://arxiv.org/abs/1802.04562} {arXiv:1802.04562 [hep-lat]} \BibitemShut
  {NoStop}%
\bibitem [{\citenamefont {Arnold}\ and\ \citenamefont
  {Zhai}(1994)}]{Arnold:1994ps}%
  \BibitemOpen
  \bibfield  {author} {\bibinfo {author} {\bibfnamefont {Peter~Brockway}\
  \bibnamefont {Arnold}}\ and\ \bibinfo {author} {\bibfnamefont {Cheng-Xing}\
  \bibnamefont {Zhai}},\ }\bibfield  {title} {\enquote {\bibinfo {title} {{The
  Three Loop Free Energy for Pure Gauge QCD}},}\ }\href {\doibase
  10.1103/PhysRevD.50.7603} {\bibfield  {journal} {\bibinfo  {journal} {Phys.
  Rev. D}\ }\textbf {\bibinfo {volume} {50}},\ \bibinfo {pages} {7603--7623}
  (\bibinfo {year} {1994})},\ \Eprint {http://arxiv.org/abs/hep-ph/9408276}
  {arXiv:hep-ph/9408276} \BibitemShut {NoStop}%
\bibitem [{\citenamefont {Braaten}\ and\ \citenamefont
  {Nieto}(1996)}]{Braaten:1995jr}%
  \BibitemOpen
  \bibfield  {author} {\bibinfo {author} {\bibfnamefont {Eric}\ \bibnamefont
  {Braaten}}\ and\ \bibinfo {author} {\bibfnamefont {Agustin}\ \bibnamefont
  {Nieto}},\ }\bibfield  {title} {\enquote {\bibinfo {title} {{Free energy of
  QCD at high temperature}},}\ }\href {\doibase 10.1103/PhysRevD.53.3421}
  {\bibfield  {journal} {\bibinfo  {journal} {Phys. Rev. D}\ }\textbf {\bibinfo
  {volume} {53}},\ \bibinfo {pages} {3421--3437} (\bibinfo {year} {1996})},\
  \Eprint {http://arxiv.org/abs/hep-ph/9510408} {arXiv:hep-ph/9510408}
  \BibitemShut {NoStop}%
\bibitem [{\citenamefont {Kajantie}\ \emph {et~al.}(2003)\citenamefont
  {Kajantie}, \citenamefont {Laine}, \citenamefont {Rummukainen},\ and\
  \citenamefont {Schroder}}]{Kajantie:2002wa}%
  \BibitemOpen
  \bibfield  {author} {\bibinfo {author} {\bibfnamefont {K.}~\bibnamefont
  {Kajantie}}, \bibinfo {author} {\bibfnamefont {M.}~\bibnamefont {Laine}},
  \bibinfo {author} {\bibfnamefont {K.}~\bibnamefont {Rummukainen}}, \ and\
  \bibinfo {author} {\bibfnamefont {Y.}~\bibnamefont {Schroder}},\ }\bibfield
  {title} {\enquote {\bibinfo {title} {{The Pressure of hot QCD up to g6
  ln(1/g)}},}\ }\href {\doibase 10.1103/PhysRevD.67.105008} {\bibfield
  {journal} {\bibinfo  {journal} {Phys. Rev. D}\ }\textbf {\bibinfo {volume}
  {67}},\ \bibinfo {pages} {105008} (\bibinfo {year} {2003})},\ \Eprint
  {http://arxiv.org/abs/hep-ph/0211321} {arXiv:hep-ph/0211321} \BibitemShut
  {NoStop}%
\bibitem [{\citenamefont {Arnold}\ and\ \citenamefont
  {Zhai}(1995)}]{Arnold:1994eb}%
  \BibitemOpen
  \bibfield  {author} {\bibinfo {author} {\bibfnamefont {Peter~Brockway}\
  \bibnamefont {Arnold}}\ and\ \bibinfo {author} {\bibfnamefont {Cheng-xing}\
  \bibnamefont {Zhai}},\ }\bibfield  {title} {\enquote {\bibinfo {title} {{The
  Three loop free energy for high temperature QED and QCD with fermions}},}\
  }\href {\doibase 10.1103/PhysRevD.51.1906} {\bibfield  {journal} {\bibinfo
  {journal} {Phys. Rev. D}\ }\textbf {\bibinfo {volume} {51}},\ \bibinfo
  {pages} {1906--1918} (\bibinfo {year} {1995})},\ \Eprint
  {http://arxiv.org/abs/hep-ph/9410360} {arXiv:hep-ph/9410360} \BibitemShut
  {NoStop}%
\bibitem [{\citenamefont {Francis}\ \emph {et~al.}(2015)\citenamefont
  {Francis}, \citenamefont {Kaczmarek}, \citenamefont {Laine}, \citenamefont
  {Neuhaus},\ and\ \citenamefont {Ohno}}]{Francis:2015lha}%
  \BibitemOpen
  \bibfield  {author} {\bibinfo {author} {\bibfnamefont {A.}~\bibnamefont
  {Francis}}, \bibinfo {author} {\bibfnamefont {O.}~\bibnamefont {Kaczmarek}},
  \bibinfo {author} {\bibfnamefont {M.}~\bibnamefont {Laine}}, \bibinfo
  {author} {\bibfnamefont {T.}~\bibnamefont {Neuhaus}}, \ and\ \bibinfo
  {author} {\bibfnamefont {H.}~\bibnamefont {Ohno}},\ }\bibfield  {title}
  {\enquote {\bibinfo {title} {{Critical Point and Scale Setting in $SU(3)$
  Plasma: an Update}},}\ }\href {\doibase 10.1103/PhysRevD.91.096002}
  {\bibfield  {journal} {\bibinfo  {journal} {Phys. Rev. D}\ }\textbf {\bibinfo
  {volume} {91}},\ \bibinfo {pages} {096002} (\bibinfo {year} {2015})},\
  \Eprint {http://arxiv.org/abs/1503.05652} {arXiv:1503.05652 [hep-lat]}
  \BibitemShut {NoStop}%
\bibitem [{\citenamefont {Burnier}\ \emph {et~al.}(2017)\citenamefont
  {Burnier}, \citenamefont {Ding}, \citenamefont {Kaczmarek}, \citenamefont
  {Kruse}, \citenamefont {Laine}, \citenamefont {Ohno},\ and\ \citenamefont
  {Sandmeyer}}]{Burnier:2017bod}%
  \BibitemOpen
  \bibfield  {author} {\bibinfo {author} {\bibfnamefont {Y.}~\bibnamefont
  {Burnier}}, \bibinfo {author} {\bibfnamefont {H.~T.}\ \bibnamefont {Ding}},
  \bibinfo {author} {\bibfnamefont {O.}~\bibnamefont {Kaczmarek}}, \bibinfo
  {author} {\bibfnamefont {A.~L.}\ \bibnamefont {Kruse}}, \bibinfo {author}
  {\bibfnamefont {M.}~\bibnamefont {Laine}}, \bibinfo {author} {\bibfnamefont
  {H.}~\bibnamefont {Ohno}}, \ and\ \bibinfo {author} {\bibfnamefont
  {H.}~\bibnamefont {Sandmeyer}},\ }\bibfield  {title} {\enquote {\bibinfo
  {title} {{Thermal Quarkonium Physics in the Pseudoscalar Channel}},}\ }\href
  {\doibase 10.1007/JHEP11(2017)206} {\bibfield  {journal} {\bibinfo  {journal}
  {JHEP}\ }\textbf {\bibinfo {volume} {11}},\ \bibinfo {pages} {206} (\bibinfo
  {year} {2017})},\ \Eprint {http://arxiv.org/abs/1709.07612} {arXiv:1709.07612
  [hep-lat]} \BibitemShut {NoStop}%
\bibitem [{\citenamefont {Bazavov}\ \emph {et~al.}(2014)\citenamefont {Bazavov}
  \emph {et~al.}}]{MILC:2013ops}%
  \BibitemOpen
  \bibfield  {author} {\bibinfo {author} {\bibfnamefont {A.}~\bibnamefont
  {Bazavov}} \emph {et~al.} (\bibinfo {collaboration} {MILC}),\ }\bibfield
  {title} {\enquote {\bibinfo {title} {{Update on the 2+1+1 Flavor QCD Equation
  of State with Hisq}},}\ }\href {\doibase 10.22323/1.187.0154} {\bibfield
  {journal} {\bibinfo  {journal} {PoS}\ }\textbf {\bibinfo {volume}
  {LATTICE2013}},\ \bibinfo {pages} {154} (\bibinfo {year} {2014})},\ \Eprint
  {http://arxiv.org/abs/1312.5011} {arXiv:1312.5011 [hep-lat]} \BibitemShut
  {NoStop}%
\bibitem [{\citenamefont {Weber}\ \emph {et~al.}(2021)\citenamefont {Weber},
  \citenamefont {Bazavov},\ and\ \citenamefont {Petreczky}}]{Weber:2021hro}%
  \BibitemOpen
  \bibfield  {author} {\bibinfo {author} {\bibfnamefont {Johannes~Heinrich}\
  \bibnamefont {Weber}}, \bibinfo {author} {\bibfnamefont {Alexei}\
  \bibnamefont {Bazavov}}, \ and\ \bibinfo {author} {\bibfnamefont {Peter}\
  \bibnamefont {Petreczky}},\ }\bibfield  {title} {\enquote {\bibinfo {title}
  {{Update on (2+1+1)-Flavor QCD Equation of State}},}\ }\href {\doibase
  10.22323/1.396.0060} {\bibfield  {journal} {\bibinfo  {journal} {PoS}\
  }\textbf {\bibinfo {volume} {LATTICE2021}},\ \bibinfo {pages} {060} (\bibinfo
  {year} {2021})},\ \Eprint {http://arxiv.org/abs/2110.03606} {arXiv:2110.03606
  [hep-lat]} \BibitemShut {NoStop}%
\bibitem [{MIL()}]{MILC}%
  \BibitemOpen
  \href@noop {} {\enquote {\bibinfo {title} {Milc qcd package},}\ }\bibinfo
  {howpublished} {\url{https://github.com/milc-qcd/milc_qcd}}\BibitemShut
  {NoStop}%
\bibitem [{ope()}]{openQCD}%
  \BibitemOpen
  \href@noop {} {\enquote {\bibinfo {title} {open-qcd package},}\ }\bibinfo
  {howpublished}
  {\url{https://luscher.web.cern.ch/luscher/openQCD/index.html}}\BibitemShut
  {NoStop}%
\end{thebibliography}%
\end{document}